\documentclass{aa}  

\usepackage{graphicx}
\usepackage{txfonts}
\usepackage{lipsum}
\usepackage{subcaption}         
\usepackage{lscape}             
\usepackage{placeins}           

\usepackage[]{hyperref}
\hypersetup{
    colorlinks=true,
    linkcolor=blue,
    filecolor=blue,      
    urlcolor=blue,
    citecolor=blue,
    }
    
\newcommand{\Msun}{M$_\odot$}
\newcommand{\cc}{cm$^{-3}$}
\defcitealias{GRAVITYCollaboration2022}{GRAVITY Collaboration 2022}

\newcommand{\EPFL}    {Institute of Physics, Laboratory for Galaxy Evolution and Spectral Modelling, EPFL, Observatoire de Sauverny, Chemin Pegasi 51, 1290 Versoix, Switzerland}
\newcommand{\IAPS}    {INAF -- Istituto di Astrofisica e Planetologia Spaziali, via Fosso del Cavaliere 100, 00133 Roma, Italy}
\newcommand{\IAA}     {Instituto de Astrofísica de Andalucía, CSIC, Glorieta de la Astronomía s/n, E-18008 Granada, Spain}
\newcommand{\ZAHITA}  {Universit\"{a}t Heidelberg, Zentrum f\"{u}r Astronomie, Institut f\"{u}r Theoretische Astrophysik, Albert-Ueberle-Straße 2, D-69120 Heidelberg, Germany}
\newcommand{\CLAP}    {Como Lake centre for AstroPhysics (CLAP), DiSAT, Universit{\`a} dell’Insubria, via Valleggio 11, 22100 Como, Italy}
\newcommand{\ENSL}    {ENS de Lyon, Université Lyon 1, CNRS, CRAL, UMR 5574, Lyon, France}
\newcommand{\MPIA}    {Max-Planck-Institut f\"{u}r Astronomie, K\"{o}nigstuhl 17, D-69117, Heidelberg, Germany}
\newcommand{\UConn}   {University of Connecticut, Storrs, CT, USA}
\newcommand{\chalmers}{Space, Earth and Environment Department, Chalmers University of Technology, SE-412 96 Gothenburg, Sweden}
\newcommand{\UniFi}   {University of Florence, Department of Physics and Astronomy, via G. Sansone 1, 50019 Sesto Fiorentino, Firenze, Italy}
\newcommand{\IWR}     {Universit\"{a}t Heidelberg, Interdisziplin\"{a}res Zentrum f\"{u}r Wissenschaftliches Rechnen, Im Neuenheimer Feld 225, 69120 Heidelberg, Germany}
\newcommand{\UniWien} {University of Vienna, Department of Astrophysics, T\"{u}rkenschanzstra{\ss}e 17, 1180 Vienna ,  Austria}
\newcommand{\UniStA}  {SUPA, School of Physics and Astronomy, University of St Andrews, North Haugh, St Andrews, KY16 9SS, UK}
\newcommand{\LJMU}    {Astrophysics Research Institute, Liverpool John Moores University, 146 Brownlow Hill, Liverpool, L3 5RF, UK}

\begin{document}

\title{Rhea-RT: Dynamical impact of Central Molecular Zone conditions on the properties of the interstellar medium and stellar feedback coupling}

\titlerunning{Inefficient feedback coupling in the CMZ}
\authorrunning{R.~Tress et al.}

\author{R.~G.~Tress      \inst{\ref{EPFL}} \thanks{robin.tress@epfl.ch} \and
        N.~Brucy         \inst{\ref{ENSL},\ref{ZAHITA}} \and                      
        P.~Girichidis    \inst{\ref{ZAHITA}           } \and            
        S.~C.~O.~Glover  \inst{\ref{ZAHITA}           } \and            
        J.~G{\"o}ller    \inst{\ref{ZAHITA}           } \and            
        M.~Hirschmann    \inst{\ref{EPFL}             } \and        
        R.~S.~Klessen       \inst{\ref{ZAHITA}, \ref{IWR}} \and                       
        T.~Peter         \inst{\ref{ZAHITA}           } \and            
        J.~Petersson     \inst{\ref{EPFL}             } \and        
        M.~C.~Sormani    \inst{\ref{CLAP}             } \and        
        L.~Armillotta    \inst{\ref{UniFi}            } \and            
        C.~D.~Battersby  \inst{\ref{UConn}            } \and            
        M.~Donati        \inst{\ref{CLAP}             } \and        
        Z.-X.~Feng       \inst{\ref{CLAP}             } \and        
        J.~D.~Henshaw    \inst{\ref{MPIA}             } \and        
        D.~R.~Lipman     \inst{\ref{UConn}            } \and            
        S.~N.~Longmore   \inst{\ref{LJMU}             } \and            
        F.~Nogueras-Lara \inst{\ref{IAA}              } \and        
        V.-M.~Pelkonen   \inst{\ref{IAPS}             } \and        
        N.~Peschken      \inst{\ref{UniFi}            } \and            
        M.~A.~Petkova    \inst{\ref{chalmers}         } \and            
        A.~Plat          \inst{\ref{EPFL}             } \and        
        S.~Reissl        \inst{\ref{ZAHITA}           } \and            
        R.~Smith         \inst{\ref{UniStA}           } \and            
        J.~D.~Soler      \inst{\ref{UniWien}          }
        }

\institute{
            \EPFL     \label{EPFL}    \and
            \ENSL     \label{ENSL}    \and
            \ZAHITA   \label{ZAHITA}  \and
            \IWR      \label{IWR}     \and
            \CLAP     \label{CLAP}    \and
            \UniFi    \label{UniFi}   \and
            \UConn    \label{UConn}   \and
            \MPIA     \label{MPIA}    \and
            \LJMU     \label{LJMU}    \and
            \IAA      \label{IAA}     \and
            \IAPS     \label{IAPS}    \and
            \chalmers \label{chalmers}\and
            \UniStA   \label{UniStA}  \and
            \UniWien  \label{UniWien}
}

\abstract{The Central Molecular Zone (CMZ) is an extreme star formation environment, characterized by higher density, higher turbulence, stronger orbital shear, and stronger magnetic field strength than the solar neighborhood. It is still debated whether classical theories of star formation hold within this extreme environment. In order to assess the impact of these different conditions on the interstellar medium (ISM) and on star formation, we present radiation magnetohydrodynamic {\sc arepo} simulations of a Milky Way-type galaxy. We set up a high-resolution ($M_{\rm cell}=20$~\Msun) region in a ring around the solar radius and in the barred region of the Galaxy to have a coherent comparison between the CMZ and the solar neighborhood. Although the high densities and strong levels of turbulence affect star formation and feedback, a key difference in the regulation of star formation between the two environments comes from the short orbital times and the strong shear in the CMZ. In particular, we highlight the role of the quick dynamical decoupling of stars and gas, which leads to periodic re-embedding events in the early lifetimes of radiating O stars. Young stellar associations are efficiently sheared apart, such that the ISM is deprived of the compounding effect of radiation and supernovae in disrupting molecular clouds. This dramatically changes the evolution of giant molecular clouds and how feedback can regulate star formation in the CMZ. Stellar feedback is no longer directly coupled to the molecular cloud from which they formed, and no strong and disruptive superbubbles can develop. The feedback instead rather acts as a background source of turbulence.}


\maketitle
\nolinenumbers

\section{Introduction} \label{sec:intro}

Galactic bars significantly affect the gas dynamics in the centers of barred galaxies and fundamentally change the picture of how the interstellar medium (ISM) and star formation (SF) evolve in the galactic environment. Bars efficiently drive large amounts of gas into inflows from the disk to the inner galaxy, where the gas generally settles into a central ring or disk-like structure \citep{Mazzuca2008,Comeron2010}. In the Milky Way, this structure is known as the Central Molecular Zone (CMZ) \citep{Morris1996, Henshaw2023} and has a diameter of approximately 600~pc. 

Recent multifrequency observations by the PHANGS (Physics at High Angular resolution in Nearby GalaxieS) collaboration \citep{Leroy2021, Emsellem2022, Lee2023} have started to uncover the physical conditions in the CMZs of many nearby spiral galaxies \citep[e.g.][]{Schinnerer2023, Sun2024}, emphasizing the complexity and variety in terms of sizes, star formation, and morphology of these regions \citep[for a review, see][]{Schinnerer2024Review}. Our own Milky Way Galaxy is barred as well, and exhibits a similar configuration, although its 3D morphology and structure are much more challenging to discern \citep{Armillotta2019,Henshaw2023,Battersby2025a,Battersby2025CMZII,Walker2025CMZIII,Lipman2025CMZIV,Sofue2025}. 

The surface density in the CMZ is higher by up to $10^3$ times than typical values in the disk. This is close to conditions experienced in the early Universe. More subtly, the bar induces a variety of other extreme and unique conditions, so that the CMZ is much more than a high surface density version of the solar neighborhood. For instance, high shear, short orbital and dynamical times, and frequent cloud collisions due to bar-induced large-scale gas inflows all occur on timescales relevant to the evolution of giant molecular clouds (GMCs) and SF \citep{Regan1997,Laine1999,Elmegreen2009,Shimizu2019,Seo2019,Hatchfield2021,Sormani2023}. It is a highly turbulent environment \citep{Rathborne2014, Rathborne2015, Federrath2016}, exhibiting an unusually steep size-linewidth relation \citep{Kauffmann2017,Tassis2022}. Finally, observations of the magnetic field in the CMZ suggest that the region is highly magnetized, with a complex magnetic field structure \citep[see, for example,][]{Crutcher1996ApJ...462L..79C,Lu2024ApJ...962...39L, Butterfield2024a, Butterfield2024b}. The abundance of prominent nonthermal filaments observed in the Galactic center are likely associated with a strong pervasive magnetic field \citep{heywood2022ApJ...925..165H, Yusef-Zadeh2022ApJ...925L..18Y}. In the dense molecular clouds, intensities of up to several milligauss have been measured, suggesting that magnetic field contributions to the dynamics of the ISM in the region are important \citep{chuss2003ApJ...599.1116C,Pillai2015ApJ...799...74P}. Strong magnetic fields can suppress star formation, and magnetic instabilities can induce turbulence and drive inflows \citep[see, for example,][]{Hennebelle2019FrASS...6....5H, Moon2023, Tress2024}. 

It is still unclear whether these effects drive a significant deviation from the classic SF scaling relations \citep[eg.][]{Sun2023} and regulating mechanisms \citep[eg. ][]{Ostriker2022}.
Because of these extreme conditions, the Galactic center is indeed a unique laboratory for testing the universality of the star formation process. The Milky Way center seems to follow the scaling law between the gas column density and the surface density of the star formation rate given by the Kennicutt-Schmidt relation \citep{Kennicutt2012}.
However, it underproduces stars by about one order of magnitude relative to what is expected by its dense ($>10^4$~\cc) gas content \citep{Longmore2013}, and some molecular clouds seem to be essentially devoid of massive star formation despite their high masses and densities \citep{Rodriguez2013, Kauffmann2017, Walker2021}. Even so, on a statistical level, the departure from these scaling relations mostly affects the more extended dense gas, while on the scale of star-forming molecular clouds, the agreement is better instead \citep{Battersby2025CMZII}. Moreover, molecular gas properties in barred galaxy centers seem distinctly offset compared to other systems with higher molecular gas surface densities and velocity dispersions \citep{Sun2020}.  

These special environmental conditions make it clear that isolated cloud simulations, while useful, are insufficient for capturing the complex feedback loops that are present in real CMZs, even if they are exposed to CMZ-like radiation fields and cosmic-ray fluxes \citep{Clark2013, Bertram2015, Cusack2025}. While achieving excellent resolution close to the protoplanetary disk scales, they fail to include the larger context, and they generally fail to reproduce observed properties such as star formation efficiencies. An improvement can be achieved by evolving similar isolated clouds in the galactic potential of the Galactic center, through which effects such as galactic shear and tidal compressions can be captured \citep{Dale2019, Kruijssen2019, Petkova2023}, but it still lacks all the interactions and complexity arising from a rich barred ecosystem, such as gas inflow and cloud collisions. 

Galactic-scale simulations, on the other hand, often lack the necessary resolution to accurately capture star formation and GMC structure, and/or they miss important physics that are needed to account for a realistic matter cycle \citep{Seo2019, Tress2020, Hatchfield2021, Tress2024, Verwilghen2024, Verwilghen2025}. Idealized setups have been developed to mimic Galactic center conditions and inflow by the bar \citep{Moon2021, Moon2022, Moon2023}. These models still overlook the self-consistent interplay between all the important components acting nonlinearly in unison to shape the gas flows in the barred region. For instance, instabilities arise in the bar lanes that cause the intermittent inflow toward the CMZ \citep{Wada2004, Sormani2017, Sormani2018}, torques between the gas and the bar regulate inflow rates, and SF in the bar lanes can modulate the accretion and the young stars can provide peculiar high-angular momentum feedback sources in the CMZ. Moreover, a coherent and systematic comparison of ISM properties and SF in the CMZ with respect to the disk using the same numerical framework is still lacking. However, it would be extremely useful for highlighting differences and understanding the regulating mechanisms in the Galactic Center. 

The aim of this paper is to use a powerful new zoom simulation that combines the Galactic-scale flows with detailed high-resolution physics to fill this gap. This allows us to study how the complex environment of the CMZ affects the evolution of a typical young star-forming region. In particular, we looked at how the effect of the dense environment, high turbulence, frequent cloud collisions, strong galactic shear, and short orbital times affect how stellar feedback couples to the surrounding ISM compared to the solar neighborhood. 
The paper is constructed as follows: we first present our numerical setup in Section \ref{sec:methods} and then compare the properties of the ISM in the solar neighborhood and the CMZ in Section \ref{sec:CompareCMZtoSolarCircle}. 
We then focus on the surrounding environment of young massive stars and the effect of their feedback on the ISM (Section~\ref{sec:env_o_stars}). The results of that analysis are discussed in Section~\ref{sec:discussion}, where we introduce a new framework of how feedback couples to the ISM in the CMZ. We discus some caveats in Section~\ref{sec:caveats} and present our conclusions in Section~\ref{sec:conclusion}.

\section{Methods} \label{sec:methods}

We designed radiation magnetohydrodynamic (MHD) simulations of the ISM of a Milky Way-like barred disk galaxy. We solve the ideal MHD equations using the code {\sc arepo} (\citealt{Springel2010}, Sec.~\ref{sec:arepo}). The gas evolves in a background barred potential fine-tuned to the Milky Way (Sec. \ref{sec:potential}) and high-resolution regions were installed in the two environments we want to compare (Galactic Center vs solar neighborhood) (Sec. \ref{sec:resolution}). The code is coupled to the relevant chemistry (Sec. \ref{sec:chemistry}), star formation (Sec. \ref{sec:StarFormation}) and stellar feedback modules such as radiative feedback and supernovae (Sec. \ref{sec:feedback}), necessary to capture GMC and SF physics. After a build-up phase (Sec.~\ref{sec:ICs}), we evolve the system for $\sim 110$~Myr. This model builds upon the Rhea simulations \citep{Goller2025} and can therefore be referred to as Rhea-RT.

\subsection{Solving the MHD equations} \label{sec:arepo}

The magnetized ISM evolves by obeying the ideal MHD equations, which we solved numerically using the moving-mesh code {\sc arepo} \citep{Springel2010,Pakmor2016,Weinberger2020}. For the divergence control of the magnetic field, we used the Powell method, which has proven to be a robust and efficient implementation, despite not enforcing the divergence-free constraint ($\nabla \cdot {\mathbf B} = 0$) to machine precision. Saturation levels of the magnetic field have been found to be relatively robust with this method, although timescales over which the magnetic field grows through dynamo action are probably not \citep{Pakmor2016}. 

Conserved quantities are discretized on an unstructured grid, and fluxes between cells are computed by solving the Riemann problem in the rest frame of the faces of the cells using the HLLD Riemann solver \citep{Miyoshi2005}. The grid is constructed on each timestep by building the Voronoi mesh of a set of mesh-generating points, which move with the direction of the flow, minimizing advection errors, improving Galilean invariance, and achieving automatic adaptive resolution with density. Cells can, however, still be de/refined by customized refinement schemes based on the problem (see Sec.~\ref{sec:resolution}). To close the set of fluid equations, we used an adiabatic equation of state for the gas with adiabatic index $\gamma = 5/3$.

\subsection{Galactic potential} \label{sec:potential}

We modeled the gravity of the different mass components of a Milky Way type galaxy by an externally imposed Galactic potential, which defines the gravitational forces felt by the gas. We used the potential introduced by \citet{Hunter2024}, which is fine-tuned to reproduce the Milky Way case and is coupled to {\sc arepo} through the galaxy modeling software {\sc agama} \citep{Vasiliev2019}. The external gravitational potential is defined by the following mass components:
\begin{equation}
\begin{split}
\Phi_{\rm tot} = & \Phi_{\rm NSC} + \Phi_{\rm NSD} + \Phi_{\rm Bar} + \\
& + \Phi_{\rm Disk} + \Delta\Phi_{\rm Spiral} + \Phi_{\rm DM},
\end{split}
\end{equation}
where $\Phi_{\rm NSC}$ is the nuclear star cluster; $\Phi_{\rm NSD}$ the nuclear stellar disk; $\Phi_{\rm Bar}$ the Galactic bar which is divided into the X-shaped boxy-peanut component and the long bar; $\Phi_{\rm Disk}$ the Galactic stellar disk, which in turn is divided into the thin and thick disk; $\Delta\Phi_{\rm Spiral}$ a rotating spiral arm perturbation; and $\Phi_{\rm DM}$ an Einasto \citep{Einasto1965,Merritt2006} dark matter halo. We refer to \citet{Hunter2024} for the details of these components and their functional dependency and parameters. 

Improvements compared to our previously employed potential \citep{Tress2020,Tress2024} are mainly in the mass distribution of the inner Galaxy, which now is in much better agreement with observations. In particular, the inner $\sim 300$~pc are now modeled in greater detail. We include now the contribution of the nuclear star cluster, which contains $\sim2.5 \times 10^7$~\Msun \citep[e.g.][]{Launhardt2002,Schodel2014a,Feldmeier2014,Chatzopoulos2015,Fritz2016,Feldmeier-Krause2017} and has an effective radius of about 4–5~pc \citep[e.g.][]{Schodel2014b,Gallego-Cano2020}, dominating the gravity within the central $\sim 20$~pc. We also include the nuclear stellar disk, which dominates the dynamics up to $300$~pc instead. This is a more extended and flattened structure with a stellar mass near $10^9$~\Msun \citep[e.g.][]{Launhardt2002,Nogueras-Lara2020,Sormani2020SimMW,Sormani2022a}, a scale length of $\sim100$~pc, and a scale height of $\sim40$~pc \citep[e.g.][]{Launhardt2002,Gallego-Cano2020,Sormani2022a}. Also the Galactic Bar model has improved substantially, which now follows the made-to-measure model from \citet{Portail2017} and adapted by \citet{Sormani2022}.  

The gravity and dynamics of the gaseous component and the newly formed stars are followed self-consistently instead by solving the Poisson equation using the n-body gravity solver of the {\sc arepo} code. The gravitational softening of the gas is adaptive and set to twice the size of a cell. Star particles, on the other hand, have a gravitational softening of $10$~pc.

The central supermassive black hole (SMBH) of the Milky Way dominates the potential only within the central parsec and is modeled in our simulations as an accreting sink particle following the approach by \citet{Petersson2025} in the Noctua suite of simulations implemented in the arepoNoctua branch of {\sc arepo}. The inclusion of this component is necessary to avoid accumulation of gas in the center which would result in spurious star formation and feedback events. The central sink particle can also be used to study gas accretion onto Sgr~A* and its resulting active galactic nucleus (AGN) activity, although these are outside of the scope of our current study. The accretion radius of the central sink is fixed at $3$~pc. Its mass is $4 \times 10^6$~M$_\odot$ which is re-initialized at the end of the pre-processing phase (Sec.~\ref{sec:ICs}). This is slightly below the estimated mass of Sgr~A* of $4.3\times 10^6$~M$_\odot$ \citepalias{GRAVITYCollaboration2022}. As the black hole mass will grow during the simulation, it will ideally reach the observed mass in the middle of the simulation time. The gravitational softening of the SMBH particle is set to $2$~pc, within the accretion radius. 

\subsection{Resolution} \label{sec:resolution}

\begin{figure}
	\includegraphics[width=\columnwidth]{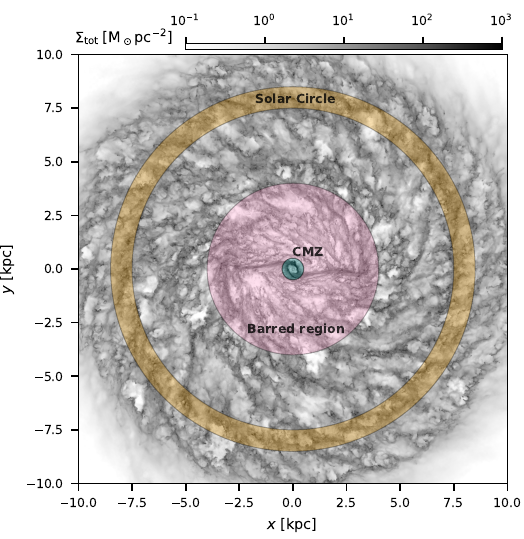}
        \caption{Face-on view of the simulated Milky Way analog. The grayscale shows the total gas column density. The highlighted areas correspond to the enhanced-resolution regions of the simulation. }
    \label{fig:HighResRegions}
\end{figure}

\begin{figure}
	\includegraphics[width=\columnwidth]{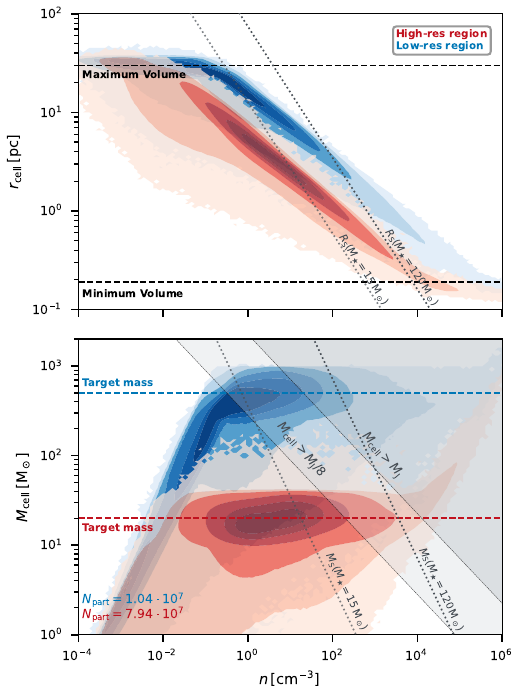}
        \caption{Resolution as a function of density. Distribution of the spherical radius (top panel) and mass (bottom panel) of the cells as a function of density. The cells in the high- and low-resolution regions are shown in red and blue, respectively.
        The contours contain 100, 99, 90, 75, 50, and 25\% of the total number of cells. The total number of cells in each region is indicated (bottom left corner in the bottom panel). We highlight the target mass in the two regions, as well as the maximum and minimum volume limits we imposed. In the bottom panel, we show with a gray overlay the region for which gas cells are more massive than one and eight times the Jeans mass. This gas is Jeans-unresolved and potentially star forming. The dotted lines indicate the demarcation lines for where the Str\"{o}mgren radius of a 15 and 120~\Msun star (which is the mass range of O-type stars) is resolved. }
    \label{fig:Resolution}
\end{figure}

Our aim was to compare ISM conditions, star formation, early cluster evolution, and feedback coupling in the CMZ with a solar neighborhood environment. To achieve this, we set up two high-resolution regions in the simulation, one region at $7.5~\mathrm{kpc} < R < 8.5$~kpc, which throughout the text we refer to as the solar circle throughout, and another region in the central $R<4$~kpc where the dynamics are controlled by the Galactic bar, which we refer to as the barred region. The central $R<500$~pc we define as the CMZ region, and our analysis focused on the ISM and SF in the CMZ region by comparing it to the solar circle. We still simulated the entire barred region at higher resolution, as the bar dynamically connects the inner part of the disk with the CMZ on timescales relevant for GMC evolution, and it is comparable to orbital timescales of gas in the CMZ. The different regions are highlighted in Fig.~\ref{fig:HighResRegions}. We set a mass resolution of $M_{\rm cell} = 20$~\Msun\ in these regions, and $M_{\rm cell} = 500$~\Msun\ for the rest of the galaxy. 

In the region $R<11$~kpc and $|z|<1$~kpc, we set a maximum volume of $V_{\rm cell, max} < 10^5$~pc$^3$, which corresponds to a spherical radius of $r_{\rm cell} \simeq 30$~pc. This is to improve the resolution in the low density region, and avoid un-physically large SN injection regions, when clustered SNe carve out low density cavities. Moreover, we restricted the volume ratio of neighboring cells to a maximum of $8$. 

When a density regime is reached where with the target mass resolution the code is unable to even resolve a single Jeans mass, it is pointless to follow the collapse further. Once this regime is reached, the gas cells were not allowed to further shrink in size as this would lead to spurious fragmentation of the unresolved collapse. To this end, we set a minimum volume for the resolution elements of $1.7$~pc$^3$ which is given by $M_{\rm cell} = M_{J}$ ie. $V_{\rm min} \propto M^3_{\rm cell}$ and corresponds to a spherical radius of $\simeq 0.2$~pc. Moreover this gas is converted into stars with an enhanced efficiency of $\epsilon_{\rm ff} = 1$ (see Section~\ref{sec:StarFormation} below). 

The mass and spatial resolution as a function of density is summarized in Fig.~\ref{fig:Resolution}. In the high-resolution region we resolve the Jeans mass with at least $8$ resolution elements up to a density of $n = 2 \times 10^2$~\cc\ and with at least one resolution element up to a density of $n = 1.3 \times 10^4$~\cc. At those densities the cold ISM has average temperatures which are above our imposed temperature floor, and so we expect the dynamics and fragmentation properties of the Jeans-resolved ISM to be properly captured. Gas at even higher densities is still present in our simulations but should be considered as unresolved and be analyzed with this caveat in mind. The dotted lines in Fig.~~\ref{fig:Resolution} indicate in which density regime H{\sc ii} regions around young O stars are resolved. The strong dependence of the Str\"{o}mgren radius with density makes it prohibitively expensive to properly resolve the early most embedded phase of every H{\sc ii} region. We discuss this in further details in the caveats section (Sec.~\ref{sec:caveats}).

\subsection{Chemistry and thermal treatment} \label{sec:chemistry}

We explicitly follow hydrogen \citep{Glover2007a, Glover2007b} and CO \citep{Nelson1997} chemistry by using a nonequilibrium time-dependent chemical network. For details of the network see \citet[][their network NL97]{Glover2012}. We opted for this rather simplified CO treatment, sacrificing accuracy in CO abundances for better performance. This is a valid compromise, as even in our high-resolution regions, the scales necessary to have converged CO abundances are mostly unresolved \citep{Joshi2019}, rendering the use of a more accurate network pointless. Still, even with this simplified network the thermodynamic behavior of the gas is excellently captured, and only minor differences are seen in the dynamics and properties of the ISM and SF at these scales compared to more expensive and accurate networks. 

Radiative heating and cooling of the gas is followed using a detailed atomic and molecular cooling function, based originally on work by \citet{Glover2010} and \citet{Glover2012} and described most recently in \citet{Mackey2019}. At low temperatures, the cooling is dominated by the ground-state fine structure transitions of C$^{+}$ and O and by rotational emission from CO molecules. The contributions of these processes are computed using the nonequilibrium abundances of these species provided by our chemical network. At higher temperatures ($T > 10^{4}$~K), permitted atomic transitions dominated. We include the effects of atomic hydrogen cooling using the nonequilibrium abundance of H$^{+}$ provided by our chemical network, but compute the cooling from He and metals in this temperature regime using the values tabulated by \citet{GF2012} that assume collisional ionization equilibrium. 

We assume the ISM to be at solar metallicity, and no chemical enrichment by feedback is considered throughout the simulation. For numerical stability reasons we set a temperature floor of $T_{\rm floor}=10$~K. The same network has been employed in similar context for previous studies \citep{Sormani2018,Tress2020,Tress2024}.

\subsection{Star formation} \label{sec:StarFormation}

\begin{figure*}
	\includegraphics[width=\textwidth]{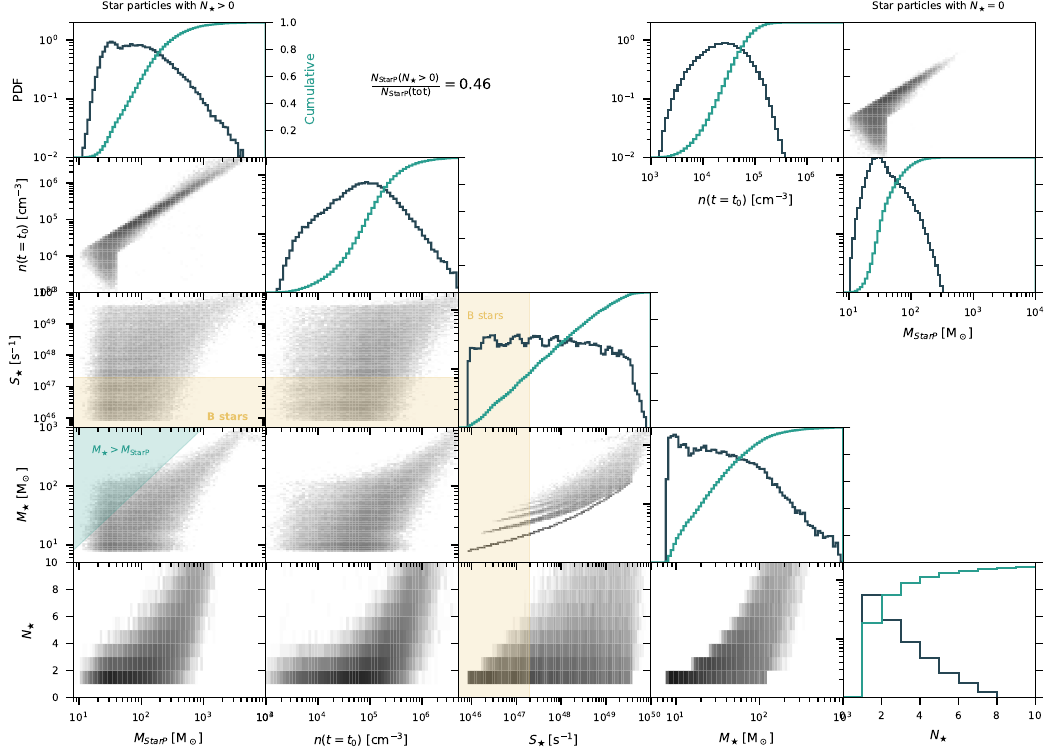}
        \caption{Properties of the star particles in the high-resolution regions. We show the distribution of the mass of the star particles (first column), the density of the gas particle that generated the star particle (second column), the ionizing radiation flux assigned to the stellar particle (third column), the total mass of stars drawn from the IMF (fourth column) and the number of stars drawn (last column). Often, no massive star is drawn from the IMF sampling. The mass and parent density distribution of those star particles is shown in the upper right. } 
    \label{fig:StarPAnalysis}
\end{figure*}

To model star formation, we employ a stochastic star particle approach, widely used in galaxy evolution research \citep[e.g.][]{Kretschmer2020, Smith2021}. Here, gas particles are stochastically converted into collisionless stellar particles based on the local properties of the gas. Contrary to sink particles, which are more commonly used in problems where local centers of collapse within the dense ISM are well resolved, these star particles interact only through gravity with the gas, and cannot accrete after they have formed \citep[see][for a comparison between sink and star particles]{Kang2025}. Given our resolution (see Sec.~\ref{sec:resolution}), the simulations are on the verge of a regime where sink particles are a valid tool. However, we still opted for the star particle approach, which is better suited for the low-resolution regions and provides a good approximation in the higher resolution ones, enabling a seamless and coherent transition between the two.

We used the implementation described in \citet{Goller2025}. Star formation is initiated when the local collapse cannot be comfortably resolved anymore based on the resolution of the simulation, i.e. when $M_{\rm cell} > M_{\rm J}/8$ (\citealt{Truelove1997}, for the equivalent MHD resolution criterion, see e.g. \citealt{Heitsch2001} or \citealt{Federrath2011}). At the current fiducial resolution we are able to resolve centers of collapse within GMCs, therefore we are more stringent in our ISM requirements to differentiate between simply dense gas and actual collapsing regions. We require that $\nabla \cdot {\mathbf v} < 0$ and that the gas cell has to be at a local minimum of the gravitational potential.     

Gas cells satisfying these criteria are stochastically converted into star particles with an efficiency per free-fall time of $\epsilon_{\rm ff} = 0.1$. As stated in Sec~\ref{sec:resolution}, at densities where $M_{\rm cell} > M_{\rm J}$ we use $\epsilon_{\rm ff} = 1$ instead.

The star particles formed by this method do not represent individual stars; rather, they are considered to be small stellar associations with masses ranging from $10-1000\,\mathrm{M}_\odot$. We therefore use the mass of the star particle to perform an Initial Mass Function (IMF) sampling and populate the particle with stars on a subgrid level. We assume a \citet[][]{Kroupa2001} IMF throughout the simulation domain. In the CMZ there are observational and theoretical hints for a top-heavy IMF \citep[e.g.][]{Hosek2019, Chabrier2024}, so an interesting future research direction would be to investigate whether the resulting difference in feedback would have a noticeable effect on gas dynamics in the Galactic Center. We use the IMF sampling scheme of \citet{Sormani2017}, which by design ensures that the IMF is always reproduced, regardless of the mass distribution of the particles to be sampled. This leads to the case where sometimes the drawn stellar mass is larger than the mass of the star particle itself. However, the drawn stars are only used to determine the feedback output from the particle, such that this prescription does not compromise mass conservation. Moreover, this only happens for a small fraction of stellar particles (Fig.~\ref{fig:StarPAnalysis}) \citep[see][for an alternative treatment to prevent such cases]{Hirai2021}. We only keep information about the mass of the O and B stars drawn from the IMF (i.e.\ stars with $M>8$~\Msun), since those are the ones that are relevant for the photoionization and supernovae (see Section~\ref{sec:feedback} below). These are the most important feedback processes at the scales we study \citep{Gatto2017,Rathjen2021}.

Throughout the paper, whenever we use the term stellar or star particle, we refer to the {\sc arepo} particle, which does not necessarily represent an individual star. The term star is instead used to indicate the individual stars drawn by the IMF sampling and assigned to each stellar particle.

\subsection{Stellar feedback} \label{sec:feedback}
Star particles are sources of feedback based on the number and mass of massive stars assigned to each of them. 

\subsubsection{Ionizing radiation feedback}

\begin{figure}
	\includegraphics[width=\columnwidth]{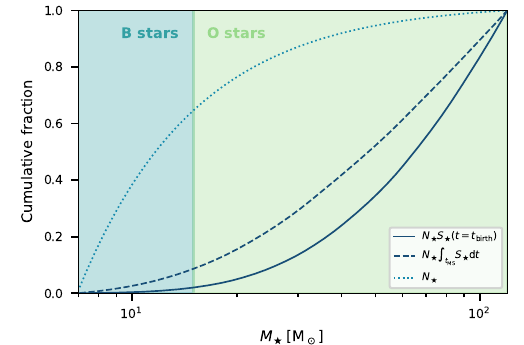}
        \caption{Contribution to the total ionizing photon budget of massive ($>8$~\Msun) stars given their mass weighted with the IMF. The dotted line shows the cumulative fraction of massive stars, the solid line the cumulative fraction of ionizing photons at birth of the star, while the dashed line shows the cumulative fraction of the ionizing photons generated over the entire life of the star as a function of mass.} 
    \label{fig:IonisationFraction}
\end{figure}

Each star particle is a potential source of ionizing radiation feedback. Each OB star drawn from the IMF and assigned to the star particle contributes to the radiation flux of the particle for the duration of its main sequence lifetime. The radiation flux is determined by assuming each star to be a black body of a given radius and surface temperature, by taking the stellar Zero Age Main Sequence (ZAMS) values from \citep{Ekstrom2012}. The radiation for each star is kept constant for the lifetime of the star. 

Under these assumptions, in Fig.~\ref{fig:IonisationFraction} we show how much stars of a given mass contribute to the total photon budget of ionizing radiation given our assumed IMF. Less than $10$\% of ionizing photons come from B stars despite their longer lifetimes and the fact that more than $60$\% of massive stars are B stars. We still consider the contribution of B stars to the ionizing radiation feedback. However, throughout the analysis we only focus on O stars and their surroundings.  

The radiation is then transported through the domain by solving the radiative transfer equations using \textsc{sweep} \citep{Peter2023}, which is a discrete ordinates method coupled to {\sc arepo}, assuming infinite speed of light approximation under steady-state-conditions. Here the equations are discretized in time, position, frequency and angular direction. For each angular direction, the cells are sorted to determine a dependency graph which is then solved using a task-based parallelization approach. This is swiftly achieved by exploiting the interconnectivity of {\sc arepo} cells through the Delaunay triangulation, which makes looping over neighboring cells trivial. Apart for a post-processing step necessary to the analysis in Section~\ref{sec:ISMPhases}, we followed a single frequency bin for photons with energy above $13.6$~eV. 

Transport sweeps are performed on every {\sc arepo} synchronization timestep, when the Delaunay triangulation for the entire domain is reconstructed. This is when photon fluxes for each cell are updated. These are then passed to the chemistry routines which compute temperature and ionization state of each gas particle. In between synchronization timesteps photon fluxes are kept constant. We forced the code to perform one synchronization timestep at least once every $4$~kyr such that the radiation is updated on a high enough cadence, and we were so able to capture variation on timescales relevant to the evolution of H{\sc ii} regions \citep[eg.][]{Maillard2021}.

\subsubsection{Diffuse interstellar radiation field}

\begin{figure}
    \centering
    \includegraphics[width=\linewidth]{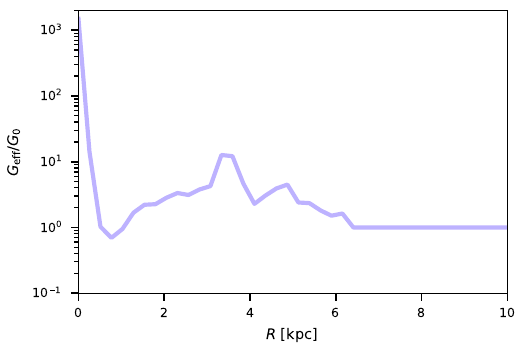}
    \caption{Background interstellar UV radiation field (photons with energy $<13.6$~eV) as a function of galactocentric radius. The value is scaled to the solar neighborhood value $G_0$ \citep{Draine1978}. The cosmic-ray ionization rate follows the same profile, scaled to the solar neighborhood value.}
    \label{fig:ISRF}
\end{figure}

The background interstellar radiation field at lower frequencies ($<13.6$~eV), which is relevant for instance for molecular hydrogen photo-dissociation and photoelectric heating, was assumed to have a constant background value depending on the galactocentric radius. Locally, for each gas cell, we computed anisotropic attenuations based on the surrounding ($<30$~pc\footnote{This is chosen to represent the average distance to a nearby O or B star in the solar neighborhood \citep{Reed2000, Maiz-Apellaniz2001}. In the CMZ this distance is most likely much smaller, so we probably overestimate shielding in this region.}) gas by computing column densities along $48$ different directions with the TreeCol algorithm \citep{Clark2012}. To estimate the radial dependence, we assumed that its profile is shaped by the younger stellar populations. Similar to what was done by \citet{Smith2023}, we computed the time-averaged star formation rate surface density as a function of galactocentric radius of a low-resolution simulation of the system, and used a rescaled version of this for the Interstellar Radiation Field (ISRF). The profile used is shown in Fig.~\ref{fig:ISRF}. The same radial dependence is then also used for the cosmic-ray ionization rate. 

\subsubsection{Supernovae} \label{sec:supernovae}

We included supernova (SN) feedback from both type II and type Ia. The former are coupled to stellar particles, and inject energy at their current location into the ISM at the end of the main sequence lifetime of their associated massive stars. These naturally generate clustered SN feedback, as stars form in associations. Type Ia supernovae instead are associated with older stellar populations and therefore distributed more randomly. We assigned them the same distribution as the density profiles of the stellar components used to generate the background potential in which the gas evolves. They were injected with a uniform distribution in time, with a rate of $1/250$~yr$^{-1}$.

The energy is inserted into the nearest neighbors to the cell closest to the location of the SN event, which are accessed by looping through the Delaunay connections of this cell \citep{Smith2018}. If the adiabatic Sedov-Taylor phase is resolved \citep{Blondin1998}, a total of $10^{51}$~erg in the form of thermal energy is injected. The hydro solver is then able to self-consistently evolve this energy into a blast wave, converting it partly to kinetic energy which will be deposited into the ISM, while still contributing to the hot $T\gtrsim10^6$~K ISM phase. On the other hand, if the Sedov-Taylor phase is not resolved, we estimate the energy of the SN remnant at the end of the momentum conserving snow-plow phase and inject this energy in the form of kinetic energy \citep{Gatto2017}. In this case, the SN event will not contribute to the hot phase, but it will deposit the right amount of kinetic energy into the ISM. This energy would have been un-physically radiated away if it would have been injected in the form of thermal energy \citep[e.g.][]{Kim2015}. Most of the SNe in the high-resolution regions (Section~\ref{sec:resolution}) are resolved, and $\lesssim 10$~\% require momentum injection.

To ensure that the energy is always injected into active cells, we modified the time-step criteria such that cells within a $50$~pc radius of the star particle have to be active when a SN event is scheduled. For stability reasons, during injection we limit the maximum temperature to $5\times10^7$~K. Except during the setup phase (Section~\ref{sec:ICs}) where we want to avoid gas depletion, each SN event only injects energy, and no mass.

\subsection{Initial conditions} \label{sec:ICs}

We used the same initial density profile as in the simulations of \citet{Tress2024}. The disk was truncated at $10$~kpc as we are not interested in the outer Galaxy in this study. The magnetic field was initialized to a purely vertical component of intensity $|{\bf B}_0| = B_z = 0.02 \,\mu$G. Turbulence and galactic rotation will bring this initial field closer to an observed configuration with saturation levels around $10\,\mu$G at the solar circle. To generate turbulence in the disk and avoid violent and coherent star formation cycles induced by the initial collapse of the smooth disk, we performed a pre-processing step. Here we reduced the lifetime of stars by a factor of five such that the feedback coupling to the ISM happens on shorter timescales and an equilibrium state can be achieved more smoothly. During this phase, the only source of feedback is in the form of SNe, and stars re-inject their mass fully into the ISM at the end of their lifetime, so that the gas is not depleted during this stage. This is performed at a low mass resolution of $500$~\Msun over a period of $100$~Myr. During this time, the background potential is kept in its axisymmetric form. The next $150$~Myr were devoted to the smooth introduction of the Galactic bar. The non-axisymmetric version of the barred potential is introduced linearly over this duration by keeping the total mass constant, leading to a steady formation of the bar lanes and CMZ \citep{Athanassoula1992}. Stellar lifetimes were slowly returned to their nominal values over the first $100$~Myr of this phase. Finally, $20$~Myr before the Bar is fully on, we increased the resolution to the reference values described in Sec.~\ref{sec:resolution} and introduced the ionization feedback. After the Bar is fully on, we turned off the $100$~\% gas re-injection, and reset the mass of the central SMBH to its value defined in Section~\ref{sec:potential} as it could have experienced excessive accretion events during this unphysical buildup phase. The simulation was then advanced in a production phase for $\sim 110$~Myr. This is enough integration time to have several independent star formation cycles and multiple rotation periods of the gas on x2 orbits in the CMZ, such that our results are statistically significant. We denote the time when the bar is fully on as $t=0$~Myr and restrict our analysis to $t>0$~Myr. During the production phase, we produced one full snapshot every $0.25$~Myr. This high cadence is necessary to follow the fast evolution of the early star formation phase, in particular in the dynamic environment of the CMZ.

\section{Comparison of the ISM properties of the CMZ and the solar circle} \label{sec:CompareCMZtoSolarCircle}

In this section, we explore in detail the properties of the simulated ISM. We emphasize the differences between the two high-resolution regions to understand and quantify the effect of the special environment of the CMZ on the ISM and SF.

\subsection{Morphology}

\begin{figure*}
	\includegraphics[width=\textwidth, trim=0.cm 0.7cm 0cm 0cm, clip]{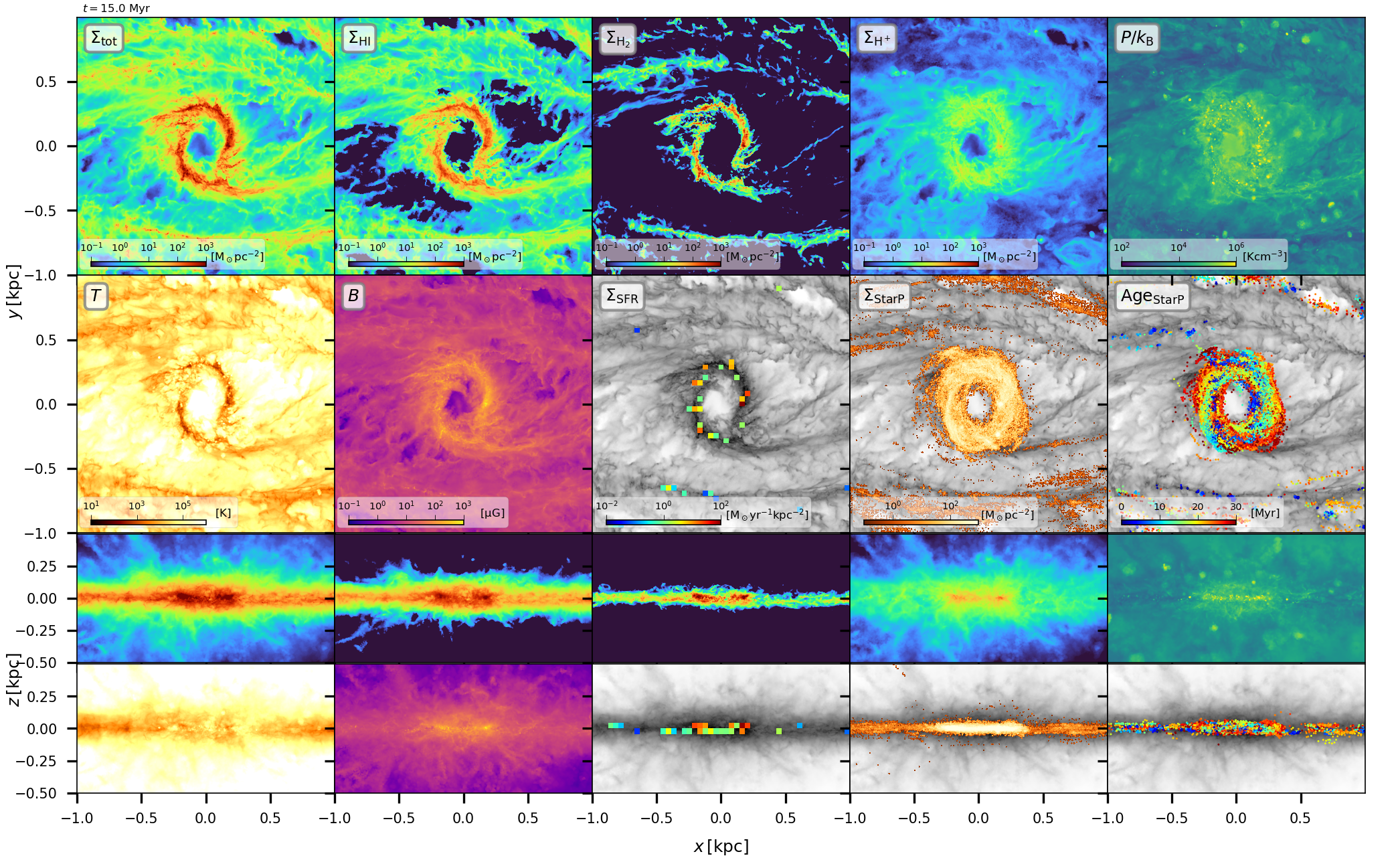}

    \vspace{0.5em}
    
	\includegraphics[width=\textwidth, trim=0.cm 0.cm 0cm 0.2cm, clip]{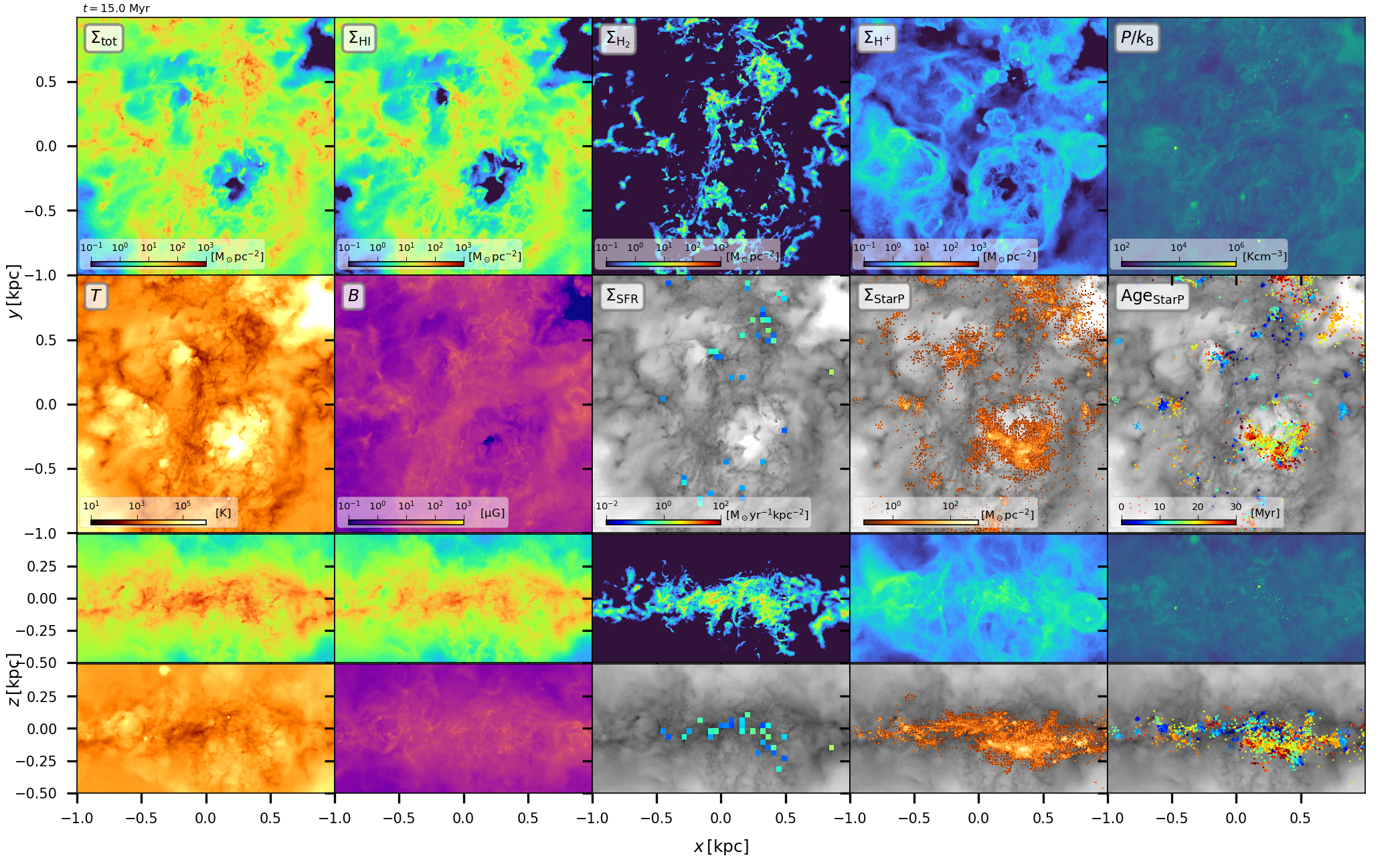}%
    \caption{Projections of the CMZ (top) and a region of the disk (bottom). The panels from left to right and top to bottom depict the total, atomic, molecular, ionized gas column densities, mass-weighted thermal pressure, temperature and magnetic field intensity averages, star formation rate surface density, stellar surface density and star particles colored by their age. These last three quantities are overlaid on the total gas column density in gray.} 
    \label{fig:Projections}
\end{figure*}

Compared to our previous modeling of the CMZ \citep{Tress2020, Tress2024}, this simulation is more advanced in terms of ISM and feedback processes included. The general large scale morphology, however, is not radically affected by this and exhibits the same features seen in simulations with simpler prescriptions. The Galactic bar dominates the gas dynamics in the central $\sim 3$~kpc, inducing strong non-axisymmetric motions. As the bar redistributes angular momentum, the gas is driven inward and flows almost radially along so-called bar lanes toward the center. The gas accumulates in a ring-like structure about $300$~pc in radius, the CMZ of this simulation. Here, torques from the bar are unable to further remove angular momentum \citep[see][for an explanation of what sets the size of the CMZ]{Sormani2024}, and other mechanisms are at play to drive gas further inward \citep{Shlosman1989, Tress2024}. The size of our simulated CMZ appears slightly larger than what is observed \citep[estimated around $100$~pc in radius][]{Kruijssen2015}, although there is some ambiguity in where exactly the CMZ ring ends \citep[section 4.3.2.2 of][]{Henshaw2023}. This is a difficult parameter to fine-tune, as it depends on the included ISM physics and feedback, as well as on resolution, so it can only be measured a posteriori once the high-resolution simulation is completed. Future iterations will account for this fact and adjust for it. The background potential also includes a four-armed spiral pattern (Section~\ref{sec:potential}), and in the disk we indeed see spiral structures develop throughout the simulation, induced by the combined effect of this spiral potential perturbation and by the rigidly rotating bar \citep[see][where the impact of the external potential on the morphology is studied in more details]{Goller2025}.

In Fig.~\ref{fig:Projections} we show a closeup of the CMZ (at the top) and a region of the same size of $2\,$kpc across in the disk, at the solar circle (bottom) at $t=15$~Myr. Without the loss of generality, we chose this simulation time as our fiducial snapshot. Many of the following figures refer to this snapshot. The ISM has settled in an equilibrium state through several SF cycles. Inspecting the bottom panel, we see dense molecular clouds forming, embedded in cold atomic envelopes. These are the formation sites of new SF events, assembling stellar clusters of different sizes and masses. The ionizing radiation associated with these young stellar clusters generate H{\sc ii} regions of different sizes and degrees of compactness. The combined feedback of these clusters is disrupting the molecular complexes from within, carving out hot expanding bubbles into the disk. The oldest clusters are found at the center of the largest bubbles, powering their expansion through SN feedback. This mechanism drives and maintains turbulence in the ISM, sustains the extended scale-height, and drives atomic and ionized outflows, which will fall back onto the disk in a fountain flow, powering future SF cycles\footnote{Movies of the simulation can be found at \href{https://doi.org/10.7910/DVN/ULCTSZ}{https://doi.org/10.7910/DVN/ULCTSZ}, which can provide a better understanding of what is described here.}.

The morphology of the CMZ here confirms the general ring-like structure identified in previous simulations. \cite{Tress2024} and \cite{Sormani2022} found that both magnetic instabilities and supernova feedback are able to transport gas from the $x_2$ ring toward the center. This model now includes both magnetic fields and stellar feedback, and indeed the same steady inflow is observed throughout the simulation. In this sense, the inner region of the CMZ is relatively devoid of gas in the initial stage of the simulation, and is gradually filled with gas at later times, similar to what is shown in \cite[][their figure 6]{Tress2024}. It is interesting to investigate how the two processes act in unison to transport gas toward Sgr A*, whether the two effects compound on each other or mutually cancel each other out. We defer this analysis to some future study. 

The CMZ structure is much more disturbed during the initial phases of the simulation, when the accretion from the bar lanes is still prominent and highly variable (top panel of Fig.~\ref{fig:SFR2}). This perturbs the CMZ gas on $x_2$ orbits, and might cause its asymmetry \citep{Sormani2018}. At later times in the simulation, the inflow is much smaller and steadier, resulting in a more symmetric appearance of the CMZ. This explains the more disk-like structure of the CMZ toward the end of the simulation. The real CMZ exhibits a highly asymmetric gas distribution \citep{Bally1988, Jenkins1994,Sormani2018,Battersby2025a}, and is likely to still experience strong accretion events from the bar lanes (\citealt{Sormani2019b, Su2025}; although lower values are reported by \citealt{Gramze2023}, their table 2), such that an early snapshot is probably closer to what is observed in the Milky Way in terms of gas distribution. 

By a visual comparison of the panels of the two environments in Fig.~\ref{fig:Projections}, we can anticipate that the CMZ has much more extreme conditions. High surface densities in the CMZ ring lead to intense star formation in a much smaller region. In the disk, for the same area, we see a handful of star-forming regions and young stellar clusters. In the CMZ, on the other hand, most of the stars form on these $x_2$ orbits, generating a ring of young stars and high stellar densities. Most of the clusters are sheared apart in the dynamic environment of the inner Galaxy, leading to streams of coeval stars (Age$_{\rm StarP}$ panel of Fig.~\ref{fig:Projections}). Only few clusters survive and stay compact ($\Sigma_{\rm StarP}$ panel). The resulting H{\sc ii} regions are generally much smaller in the CMZ, but the high density of young stars also substantially increases the diffuse contribution of ionized gas compared to the disk. At the solar circle, stellar clusters are able to drive extended superbubbles that have sizes comparable to the entire CMZ. Instead, in the Galactic Center we do see feedback-driven bubbles in the ISM, but of considerably smaller size, and which are not able to significantly affect the overall structure of the CMZ. 

In the CMZ a lot of complexity is happening in a very small region. It is extremely important for a simulation to be able to capture these complex dynamics. Inability to do so might result in widely over estimation of inflow rates \citep[e.g. appendix B of][]{Tress2024}, BH accretion, star formation rates and major differences in size and morphology of the CMZ.  

\subsection{Star formation}

\begin{figure}
    \centering
    \includegraphics[width=\linewidth]{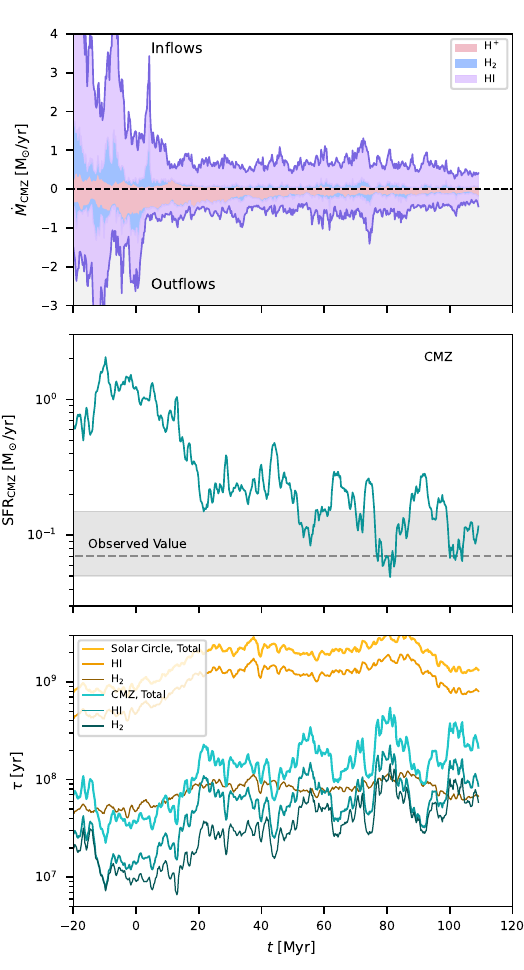}
    \caption{Gas inflow/outflow to the CMZ (top), star formation rate in the CMZ (middle), and depletion times for the CMZ and solar circle (bottom) as a function of simulation time. Inflows and outflows are shown separately and the different gas phases stacked in different color to show the contribution of each phase to the total gas inflow/outflow shown as the solid purple line. Depletion times for the atomic, molecular and total gas are shown separately. The observed SFR value for the CMZ and its uncertainty is shown as the dashed line and shaded region in the middle panel.}
    \label{fig:SFR2}
\end{figure}

\begin{figure}
    \centering
    \includegraphics[width=\linewidth]{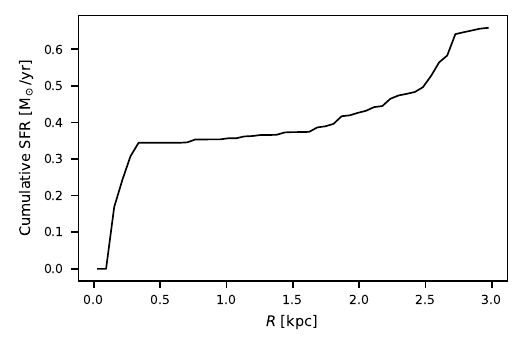}
    \caption{Cumulative star formation as a function of galactocentric radius at $t=15$~Myr. }
    \label{fig:SFR1}
\end{figure}

\begin{figure}
    \centering
    \includegraphics[width=\linewidth]{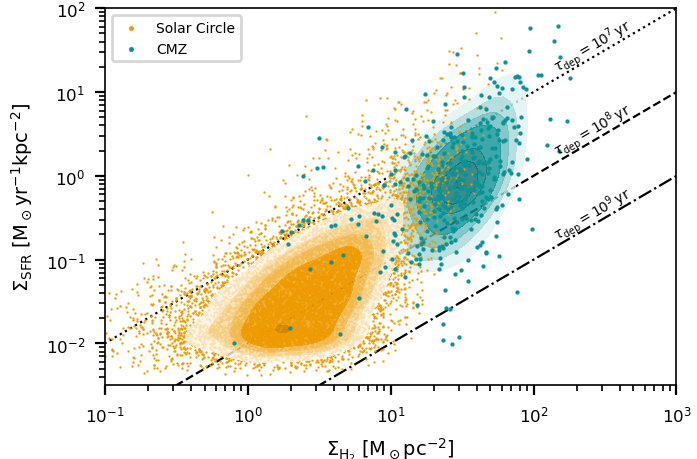}
    \caption{Star formation rate surface density as a function of molecular gas surface density. The values are computed over a $100$~pc sized bins and shown separately in different colors for the CMZ (teal) and the solar circle (orange). The values are taken for several snapshots throughout the simulation. The colored area shows the kernel density estimation of the distributions and the black lines show lines of constant depletion time.}
    \label{fig:SK}
\end{figure}

In Fig.~\ref{fig:SFR2} (middle panel) we show the star formation rate (SFR) in the CMZ region as a function of time. Compared to the simulations of \citet{Tress2020} where only SN feedback was considered, the SFR is more variable in time \citep[][their figure 2]{Sormani2020SimMW}. Still, the variation is contained within one order of magnitude in a time span of 100 Myr and the period of intense star formation ($\gtrsim 0.3$~\Msun~yr$^{-1}$) is correlated to the initial phase of intenser inflow rates (top panel of Fig.~\ref{fig:SFR2}), suggesting that in our simulation,  the SFR of the CMZ is regulated by the inflow rate onto the central region \citep[like in the simulations of][]{Moon2022}. In contrast with the simulation of \citet{Armillotta2019}, we do not see episodes where the feedback is able to significantly disrupt the gas in the CMZ after a period of intense star formation, completely shutting star formation. Our simulations, therefore, favor the inflow regulated star formation scenario over the episodic feedback regulated one. The second part of our study (Sections~\ref{sec:env_o_stars} and \ref{sec:discussion}) is dedicated to understand how feedback is coupled to the ISM in the CMZ and provides a possible explanation of why star formation is inflow-regulated there. At later times in the simulation the SFR is comparable to observed values. Most of the star formation is contained in the CMZ ring, and only very few stars form in the region between $0.5$-$2.5$~kpc where the gas dynamics are dominated by the bar lanes (Fig.~\ref{fig:SFR1}).

In the bottom panel of Fig.~\ref{fig:SFR2} we compare depletion times ($\tau = M_{\rm gas}/SFR$) in the two regions of interest. The HI depletion times is higher by an order of magnitude in the solar neighborhood than  in the CMZ, while the H$_2$ depletion times are similar in both regions. 
This suggests that the widely different conditions in the CMZ makes the transition from atomic to molecular gas much more efficient.  At first sight, it seems that the star formation process in the dense gas may proceed in a similar fashion in both region. 
However, subtle differences exist. This is emphasized in Fig.~\ref{fig:SK} where we show star formation rate surface densities ($\Sigma_{\rm SFR}$) as a function of molecular gas surface densities ($\Sigma_{\rm H_2}$) in $100$~pc bins. CMZ bins have generally higher $\Sigma_{\rm H_2}$ values, but have lower values of $\Sigma_{\rm SFR}$ compared to bins at the solar circle at similar surface densities. This indicates that in the CMZ the dense molecular gas is less efficiently converted into stars as if it would have been located at the solar circle with the same density.

\subsection{ISM phases} \label{sec:ISMPhases}

\begin{figure*}
	\includegraphics[width=\textwidth]{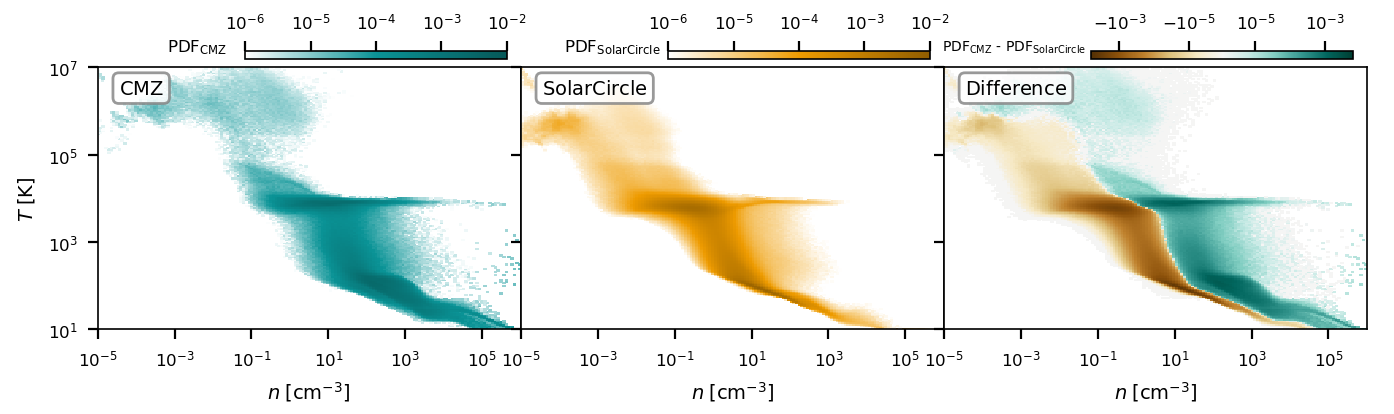}
        \caption{Density-temperature phase diagram of the gas in the simulation at $t=15$~Myr. The colors indicate the gas mass fraction in the given $(n-T)$ bin normalized by the bin size (${\rm d} M_{\rm gas}(n,T)/M_{\rm gas, tot} / {\rm d}(\log_{10} T) / {\rm d}(\log_{10} n)$). The gas in the CMZ is shown in the left panel, the gas at the solar circle in the middle panel, and the difference on the right.} 
    \label{fig:nTPDF}
\end{figure*}

\begin{figure}
	\includegraphics[width=\columnwidth]{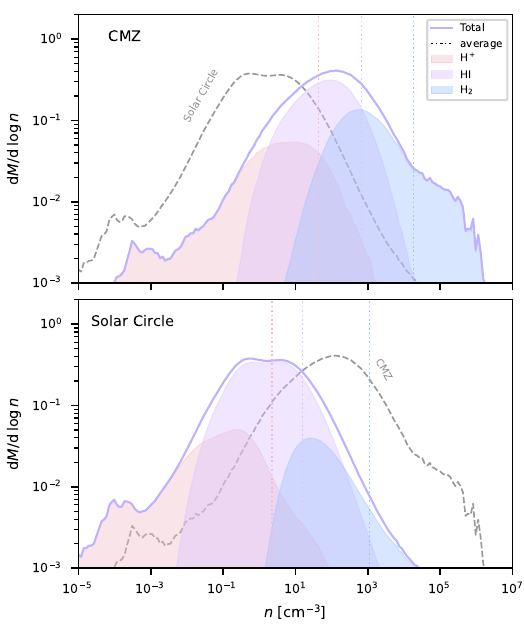}
        \caption{Number density probability distribution function for the CMZ (top) and the solar circle (bottom) at $t=15$~Myr. The contribution of the different ISM phases is shown in different colors. Dotted vertical lines indicate average density values for the different phases.} 
    \label{fig:DensityPDF}
\end{figure}

\begin{figure}
	\includegraphics[width=\columnwidth]{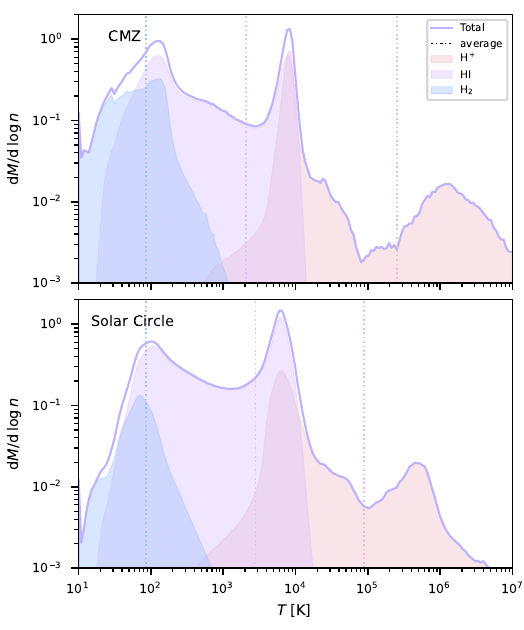}
        \caption{Same as Fig.~\ref{fig:DensityPDF}, but for the temperature.} 
    \label{fig:TPDF}
\end{figure}

\begin{figure}
	\includegraphics[width=\columnwidth]{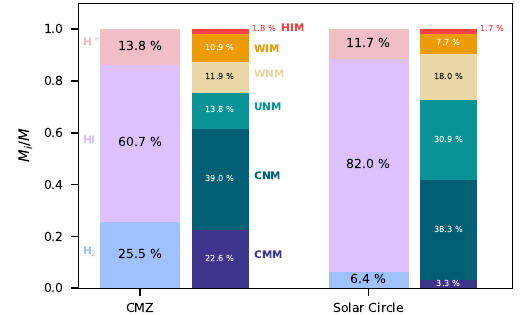}
        \caption{Mass fraction in different ISM phases for the two high-resolution regions at $t=15$~Myr. } 
    \label{fig:ISMPhases}
\end{figure}

Before comparing the different ISM phases in the two regions of interest, we first did a post-processing radiative transfer step to improve the ionization state and temperature of H{\sc ii} regions and better identify the photo-ionized gas phase. During this stage we only allow the radiation transfer and the thermochemistry to evolve in time performing 10 radiation sweeps over $10$~kyr. Contrary to the production run, in this step we include 3 frequency bins for the ionizing radiation ($(13.6-15.4)$~eV, $(15.4-24.6)$~eV, $(24.6-\infty)$~eV)\footnote{This was only implemented later, after the main simulation was already completed. This will slightly change temperatures in H{\sc ii} regions, but only has a small impact on the dynamics.} instead of a single $>13.6$~eV one. Phases other than the photoionized gas are largely unaffected by this.    

We show the mass weighted density-temperature phase space distribution in Fig.~\ref{fig:nTPDF} and the mass weighted probability density function (pdf) of the density (Fig.~\ref{fig:DensityPDF}) and the temperature (Fig.~\ref{fig:TPDF}) in the two regions. The density pdf in both cases roughly has a log-normal profile with a high-density power-law tail in molecular gas \citep[e.g.][]{Brunt2015}, and is shifted to higher densities in the case of the CMZ compared to the solar circle. The total density pdf in the CMZ peaks at $n\simeq 1.3 \times 10^2$~\cc, and at the solar circle at $n\simeq 6 \times10^{-1}$~\cc. The shift toward higher density is present even for the distribution of ionized, atomic and molecular gas taken individually.

The ISM is then divided into the following phases \citep{Draine2011Book}:
\begin{itemize}
    \item[-] Cold Molecular Medium (CMM) when $x_{\rm H_2} > 0.25$;
    \item[-] Cold Neutral Medium (CNM) when $x_{\rm H} > 0.5$ and $T<500$~K;
    \item[-] Unstable Neutral Medium (UNM) when $x_{\rm H} > 0.5$ and $500\, {\rm K}<T<6000$~K;
    \item[-] Warm Neutral Medium (WNM) for $x_{\rm H} > 0.5$ and $T>6000$~K;
    \item[-] Warm Ionized Medium (WIM) when $T < 3.5\times 10^4$~K and $x_{\rm H^+} > 0.5$;
    \item[-] Hot Ionized Medium (HIM) when $T > 3.5\times 10^4$~K.
\end{itemize}

In Fig.~\ref{fig:ISMPhases} we show the mass fraction in these different phases for the two regions for a snapshot of the simulation at $t=15$~Myr. A larger fraction of the ISM is molecular in the CMZ compared to the solar circle, and the majority of the gas is at low temperatures, either in the CNM or CMM. Still, a significant amount of gas is actually in the form of atomic hydrogen. The WIM is gas which is photo-ionized, either in the form of diffuse ionized gas, or in H{\sc ii} regions, while the HIM is mostly generated by SNe. Both the CMZ and the solar circle have comparable fractions of gas in the ionized phases (WIM and HIM) despite the much higher star formation density in the CMZ which accounts for higher feedback.

\subsection{Magnetic field}

\begin{figure}
	\includegraphics[width=\columnwidth]{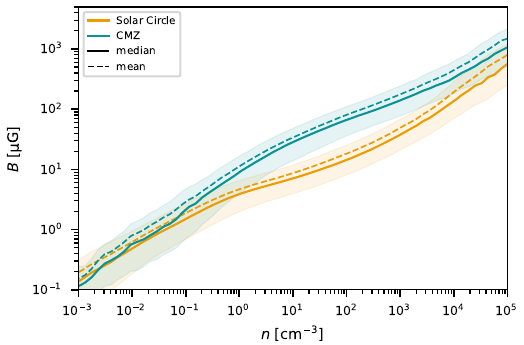}
        \caption{Average (dashed line) and median (solid line) magnetic field intensity as a function of density for gas in the CMZ (teal) and at the solar circle (orange) for the snapshot at $t = 15$~Myr. The shaded area around the mean indicates $\pm\sigma$ dispersion.} 
    \label{fig:MagneticField}
\end{figure}

Due to flux freezing, magnetic fields are generally higher in regions of higher density. As the average density in the Galactic Center is much larger than the solar circle, it is natural to expect higher magnetic field averages as well. We measured magnetic fields in the high-resolution regions averaged over the simulation time when the Bar is fully formed. The volume-weighted average magnetic field in the CMZ is $11.9 \pm 1.4$~$\mu$G compared to $2.7 \pm 0.3$~$\mu$G for the solar circle. If we compute the mass-weighted average instead, we get values of $131.4 \pm 33.6$~$\mu$G for the CMZ, and $10.2 \pm 2.2$~$\mu$G for the solar circle. 

But magnetic fields in the CMZ are not only higher due to a larger fraction of gas being at higher densities. Even in the same density regime, magnetic fields seem to be higher in the CMZ environment. This can be seen in Fig.~\ref{fig:MagneticField}, where the average magnetic field intensity as a function of density is shown for the solar circle and the CMZ in the simulation. As the gas in the Galactic Center is more turbulent compared to the disk (see Sect.~\ref{sec:Turbulence}), and orbital times are much shorter, we can have a more efficient action of small-scale dynamo and large-scale mean field dynamo, increasing the magnetic field. This can also be visualized in terms of energy equipartition, which implies higher magnetic fields for regions of higher kinetic energy in turbulent motions. Both environments largely follow a relation $B \propto n^{0.5}$, consistent with what is reported by \citet{Whitworth2025} for observations and simulations \cite[see also][]{Crutcher2012}. The difference in the exponent between the solar circle and the CMZ for intermediate density regimes might be explained by different strengths and modality of the turbulent driving mechanism \citep{Sur2012} in the two regions.

The difference in $B$ as a function of density for the two regions might not seem significant, but folding the information of Fig.~\ref{fig:MagneticField} with the density distribution in Fig.~\ref{fig:DensityPDF}, we notice that most of the gas in the CMZ sits at densities where the difference is largest. Moreover, the value of $B$ in the CMZ consistently lies above the solar circle value throughout the different density regimes, which accounts for a significant total energy difference in magnetic fields for the two regions.

Our results of a stronger field in the CMZ than in the solar neighborhood are in line with observational findings by 
\cite{Akshaya2023MNRAS,Akshaya2024MNRAS}. Using multiwavelength dust polarization studies, they found plane-of-sky strengths of $2$-$7~\mathrm{mG}$ in the circumnuclear disk and mini-spiral, with pronounced spatial variations. 
Their finding of particularly weak fields along the Eastern Arm of the mini-spiral, where the Alfven Mach number exceeds unity, resonates with our picture of local variations in feedback coupling and turbulence driving. 

\subsection{Turbulence}\label{sec:Turbulence}

\begin{figure}
	\includegraphics[width=\columnwidth]{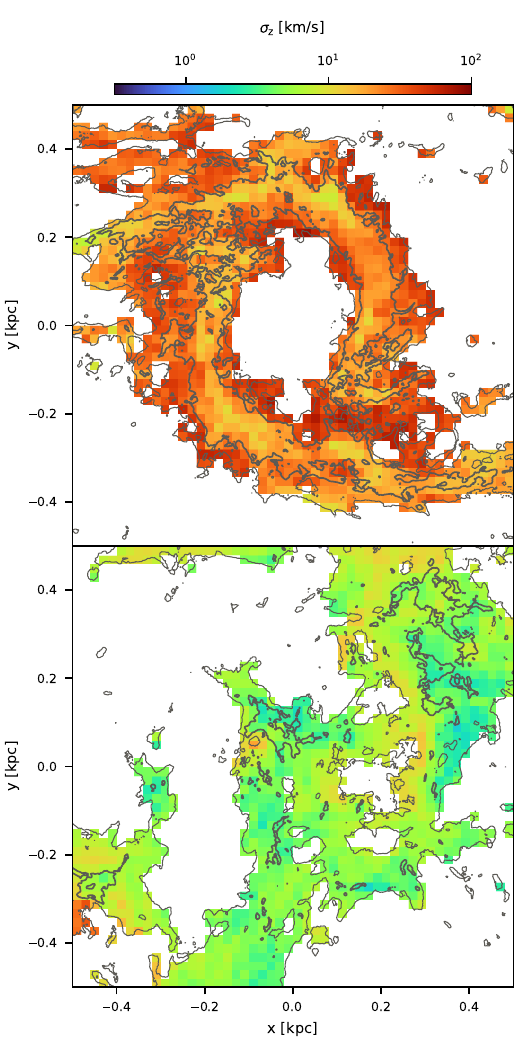}
    \caption{Velocity dispersion ($\sigma_z$) map for the CMZ and a section of the disk at $t=15$~Myr. Values are obtained by inferring the mass-weighted velocity dispersion of cells in $20$~pc bins. The contours show column densities of $[10, 50]$~\cc. Lower density bins are excluded for clarity. } 
    \label{fig:sigmaz_map}
\end{figure}

\begin{figure}
	\includegraphics[width=\columnwidth]{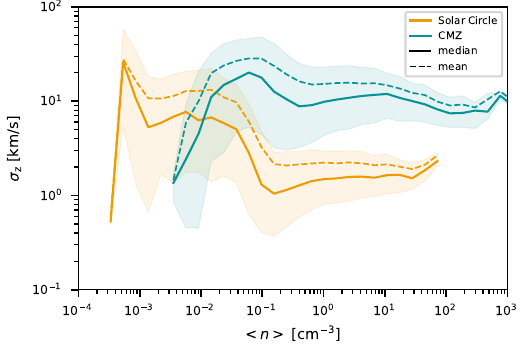}
    \caption{Average (dashed line) and median (solid line) z component of the velocity dispersion as a function of density for gas in the CMZ (teal) and at the solar circle (orange) at $t=15$~Myr. The shaded area around the mean indicates the standard deviation.} 
    \label{fig:sigmaz_vs_rho}
\end{figure}

To estimate the turbulence in our simulated ISM, we examined the velocity dispersion of the gas. In particular, in the Galactic Center, the velocity dispersion can be quite asymmetric as the in-plane motions are dominated by the strong shear induced by the differential galactic rotation. To avoid accounting for these coherent shearing motions, we only computed the $z$ component of the velocity dispersion, which instead is not directly affected by this. 

In Fig.~\ref{fig:sigmaz_map} we show a $\sigma_z$ map of the CMZ and an equally sized region at the solar circle at $t=15$~Myr. The velocity dispersion is computed in bins $20$~pc in size. The higher degree of turbulence becomes evident. To quantify this further, we provide in Fig.~\ref{fig:sigmaz_vs_rho} the (spatially-) averaged value of $\sigma_z$ as a function of density for the two regions of interest. Especially for higher density regimes the gas in the CMZ is significantly more turbulent with velocity dispersions averaging around $\sim 9$~km/s compared to values around $2$~km/s for the solar circle. The reason for this is more energetic stellar feedback due to higher star formation rate surface densities in the Galactic center, stronger galactic shear cascading down to turbulent motions, more frequent cloud collisions in the dense and dynamic environment, violent gas accretion from the dust lanes and magnetic instabilities whose nonlinear regime manifests as turbulence. In addition, the feedback of wandering young massive stars could also be an important source of turbulence in the CMZ \citep{Tassis2022}.

\subsection{Summary: Differences between the solar circle and the CMZ}

In Table~\ref{tab:ComparisonTable} we report properties of the ISM in the CMZ and solar circle regions, time-averaged over $0<t<110$~Myr. 

To retrieve total, atomic, molecular, ionized gas column densities and SFR surface densities, we first computed the profile of these properties as a function of galactocentric radius for each snapshot (using a $30$~pc radial bin). The solar circle value for each snapshot is then taken by averaging these profiles over $7.5\,{\rm kp}c<R<8.5$~kpc. In the case of the CMZ, we took the maximum value of the profile within $R<0.5$~kpc instead. This is to get a better estimate of the actual column density encountered in the CMZ ring, as its size varies slightly over the course of the simulation, and averaging over the entire $R<0.5$~kpc would arbitrarily dilute these values. 

Velocity dispersion and magnetic field intensities for each snapshot are obtained by computing the mass-weighted average of these values over each region. 

To infer the SN-rate surface density, we logged the position of each SN over $5$~Myr bins. We account for the galactic rotation by rotating the SN positions by $\theta = \Omega(R)(t-t_{\rm 0, bin})$, where $\Omega(R)=v_{\rm rot}/R$ is the angular rotation velocity of the galaxy. For the SNe in the solar circle we take $\Omega(R=8\,\rm kpc)$ and $\Omega(R=0.2\,\rm kpc)$ for the CMZ. This ensures that the SN-rate surface density does not artificially decline simply because the parent cluster moves along its Galactic orbit. However, shearing forces, disrupting the cluster, will decrease this values. The reported value in Table~\ref{tab:ComparisonTable} is then obtained by time-averaging all the values obtained over these 5~Myr time intervals. 

Finally the mean ionization parameter was obtained by computing a photon number weighted average over each region of the ionization parameter $U_i = \phi_i / (n_{\rm H, tot, i}c)$ where $\phi$ is the photon flux and $n_{\rm H,tot}$ the nucleon number density of each cell. Again, the reported value is then the time-average of these values.

\begin{table}
\centering
\caption{Comparison of different properties of the solar circle and the CMZ.}
\begin{tabular}{c c c c c}
 \hline
 \hline
 Physical quantity & Solar circle & CMZ \\
 \hline
 $\Sigma_{\rm tot}$    $[\rm M_\odot pc^{-2}]$            & $11.2  \pm  1.5$  &  $129  \pm 15$      \\
 $\Sigma_{\rm HI} $    $[\rm M_\odot pc^{-2}]$            & $0.49  \pm  0.13$ &  $42.6 \pm 8.6$     \\
 $\Sigma_{\rm H_2}$    $[\rm M_\odot pc^{-2}]$            & $6.68  \pm  0.76$ &  $51.9 \pm 7.5$     \\
 $\Sigma_{\rm H^+}$    $[\rm M_\odot pc^{-2}]$            & $0.81  \pm  0.22$ &  $4.85 \pm 2.31$    \\
 $\Sigma_{\rm SFR}$    $[\rm M_\odot yr^{-1}kpc^{-2}]$    & $0.0065\pm0.0035$ &  $2.92 \pm 2.78$ \\
 
 $\log_{10}\tau_{\rm dep, \, tot}$$[\rm yr]$ & $9.28 \pm 0.13$   &  $8.16 \pm 0.27$                        \\
 $\log_{10}\tau_{\rm dep, \, HI}$ $[\rm yr]$ & $9.06 \pm 0.13$   &  $7.78 \pm 0.28$                        \\
 $\log_{10}\tau_{\rm dep, \, H_2}$$[\rm yr]$ & $7.91 \pm 0.09$   &  $7.54 \pm 0.30$                        \\
 
 $\sigma_z$     $[\rm km/s]$                    & $2.09  \pm 0.36$  & $8.57 \pm  1.98$                \\
 $B$              $[\rm \mu G]$                  & $10.2 \pm 2.2 $   & $131  \pm 33$               \\
 $f_{\rm H_2}$                     & $0.06 \pm 0.01$   & $0.34 \pm 0.06$                          \\
 $f_{\rm HI}$                      & $0.84 \pm 0.02$   & $0.58 \pm 0.04$                               \\
 $f_{\rm H^+}$                     & $0.10 \pm 0.01$   & $0.07 \pm 0.03$                              \\
 $\log_{10}\Sigma_{\rm SN-rate}$   $[\rm yr^{-1} pc^{-2}]$      & $-2.25 \pm  0.32$ & $-1.62 \pm  0.32$  \\
 $\log_{10} U$                     & $-2.68 \pm 0.13$  & $-2.47  \pm  0.29$         \\
 
 \hline
\end{tabular}
\tablefoot{The quantities are time-averaged over the course of the simulation. From top to bottom, the properties are total gas, atomic, molecular and ionized gas column densities, star formation rate surface density, depletion times for the total gas, the HI and the H$_2$, z component of the velocity dispersion, magnetic field intensity, molecular, atomic and ionized gas fractions, SN-rate surface density, and finally, ionization parameters. In the text we describe how these quantities are computed. }
\label{tab:ComparisonTable}
\end{table}
    
\section{Environment around O stars}
\label{sec:env_o_stars}

After highlighting the differences between the two environments, we focus now on the question whether these differences can affect stellar feedback coupling with the surrounding gas. To investigate this, we more closely inspect the environment around young O stars. These are the main source of feedback and are central for the early disruption of the parental clouds. How effective are they in doing so in our simulation and in particular in the two different environments?

\subsection{Life and death of a GMC at the solar circle}

\begin{figure}
	\includegraphics[width=\columnwidth]{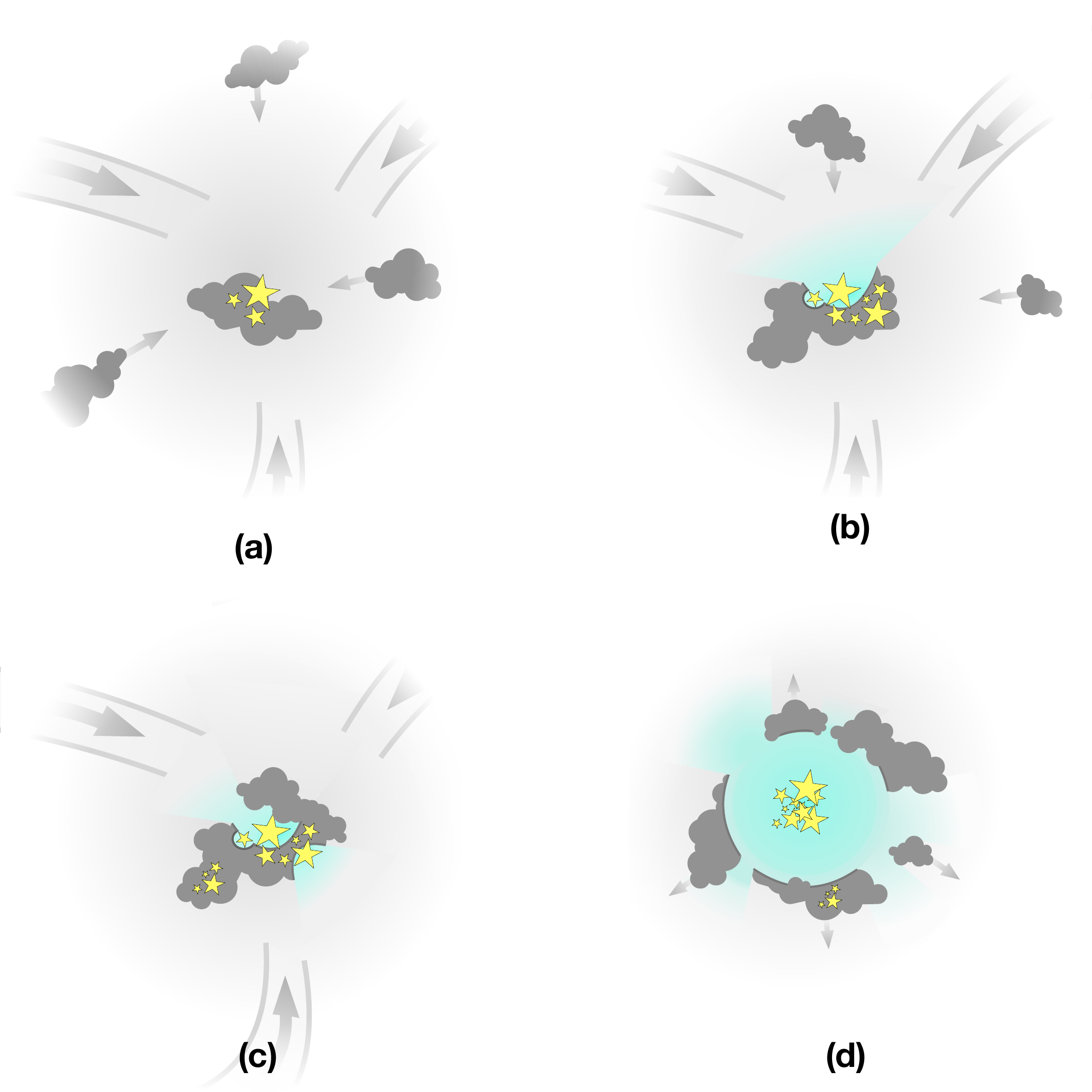}
        \caption{Schematic representation of the assembly and disruption of a GMC. Stars form as the cloud is assembling (panel a,b and c), until their combined feedback is strong enough to disrupt the cloud from within (panel d).} 
    \label{fig:SkechDisk}
\end{figure}

\begin{figure*}
    \centering
	\includegraphics[width=\textwidth, trim=0.cm 0.7cm 0cm 0cm, clip]{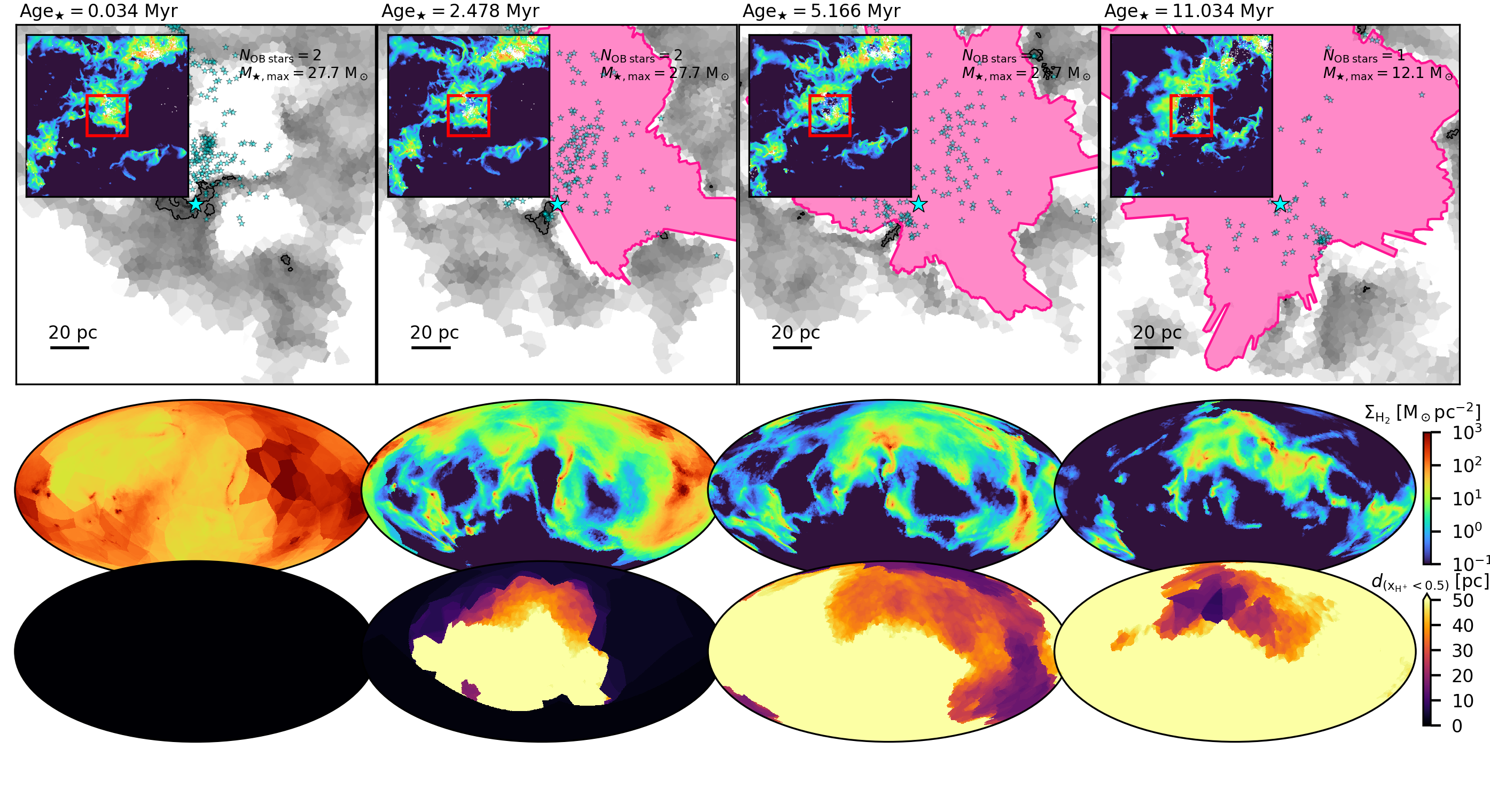}
    
    \vspace{4em}
    
    \includegraphics[width=\textwidth, trim=0.cm 0.7cm 0cm 0cm, clip]{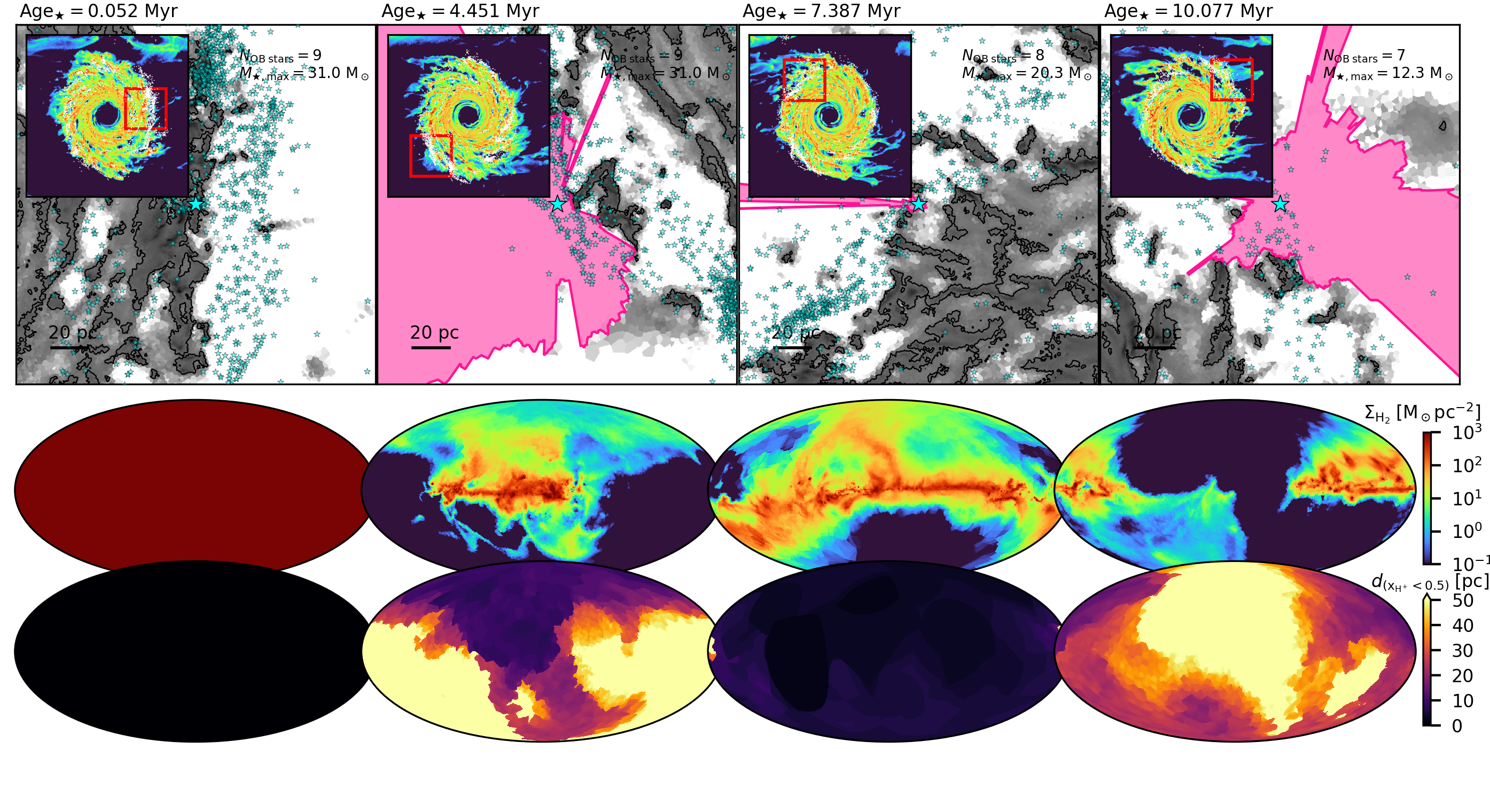}
    \caption{Time evolution of the environment surrounding an O star at the solar circle (top three rows) and in the CMZ (bottom three rows). In each panel, the first row shows the H$_2$ density slice surrounding the selected star, which is centered in the box. Iso-density contours are shown at a density of $100$~\cc. The inlay panels shows the H$_2$ column density of the larger-scale context. Surrounding O stars are shown as white dots in the inlay and turquoise stars in the main panel. We highlight the age of the selected star particle, the total number of OB stars assigned to it, and the mass of the most massive one among those. The magenta area is the region visible by the ionizing radiation of the star and which the star contributes to keeping ionized, i.e.\ the region of space for which rays connecting to the star only pass through ionized gas ($x_{\rm H^+} > 0.5$). The second and third rows show the H$_2$ column density and size of the H{\sc ii} region, as seen by an observer at the position of the star. The second row is the Mollweide H$_2$ column density projection. The third row shows the size of the H{\sc ii} region surrounding the star. As defined here, the size is the distance in each direction at which the fractional ionization of the gas first drops below 50\% (i.e.\
    $x_{\rm H^+} < 0.5$). Note that the panels are not equally spaced in time, but instead a few salient snapshots in the evolution of the star are selected. } 
    \label{fig:HIIevo}
\end{figure*}

\begin{figure*}
    \centering
    \includegraphics[width=\textwidth]{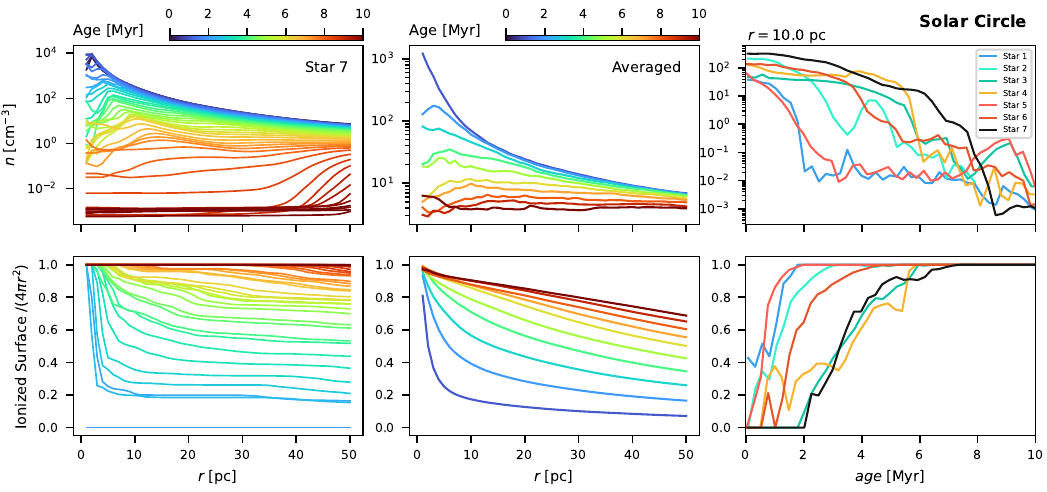}
    \includegraphics[width=\textwidth]{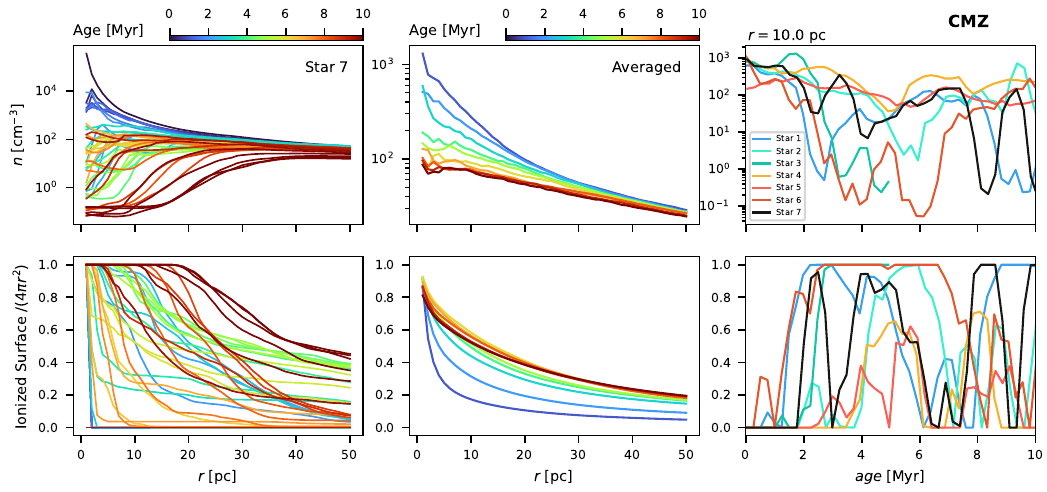}
    \caption{Density and ionization profile around O stars. The top two rows are the case for the solar circle, and the bottom two rows for the CMZ. For each row, the first column is an example for one selected star, while the second column is the average over all stellar particles in the region (solar circle or CMZ) that contain at least one O star. For each region, the first row shows the density as a function of distance from the O stars, averaged over spherical shells. The second row depicts the fraction of the surface that the O stars are helping to keep ionized, i.e. the fraction of a surface of a given radius centered on the star for which a ray passes only through ionized gas ($x_{\rm H^+} > 0.5$). These quantities are shown at different ages of the star particles. The last column shows the time evolution of the environment around a selection of seven O stars at a radius of 10 pc.}
    \label{fig:HIIRegion}
\end{figure*}

A classical star formation site in a dynamic environment would follow a time evolution as illustrated schematically in Fig.~\ref{fig:SkechDisk} \citep[e.g.][]{Geen2016, Grudic2021, Kim2021, Brucy2025}. Stars form in the dense parts of a GMC, which itself forms due to several instabilities in the gas-rich disk of a Milky Way-type galaxy. As soon as stars start to form, massive stars will influence the surrounding ISM by strong ionization feedback and stellar winds\footnote{stellar winds are not included in this simulation.}. However, a single star is unlikely to be able to disperse the entire dense surrounding molecular gas. Moreover, in a dynamic and interacting environment, more gas might be flowing toward the star formation site and more molecular gas might be condensing from the atomic phase, which leads to more star formation. Stars, therefore, never form alone, and the formation phase in a GMC complex is not instantaneous, but might be happening over timescales of a few Myr. In this initial phase, stars might be partially or totally embedded in the dense cloud and the H{\sc ii} region is ionization bounded (no ionizing radiation is leaving this region). Sooner or later, however, enough stars might have formed such that the combined feedback is able to disperse the surrounding gas, and shut further star formation off in the region. The cluster is now fully exposed, and the H{\sc ii} region becomes density bounded, allowing a large fraction of the ionizing photons to escape. 

In Fig.~\ref{fig:HIIevo} (top) we show the evolution of an example star-forming region at the solar circle as experienced from the position of an O star associated with the region. It essentially follows the story sketched out in Fig.~\ref{fig:SkechDisk}, as the GMC structure is slowly being disrupted from within by stellar feedback. 
More quantitatively, in Fig.~\ref{fig:HIIRegion} (top left panel), we look at the spherically-averaged density profile around  this O star as a function of time. The progressive effect of internal feedback disrupting the cloud around the star and expanding the ionized bubble becomes evident. The left panel of the second row of Fig.~\ref{fig:HIIRegion} shows the structure of the H{\sc ii} bubble around the star with time. We see that the new-born star is fully embedded and the H{\sc ii} region is initially small, with size $\lesssim 10$~pc. It remains ionization bounded until $1$-$2$~Myr when the radiation is able to carve out small windows and a small fraction of photons is able to escape to large distances. Slowly, the H{\sc ii} region tends to widen, as the windows' size increases and a progressively larger fraction of photons escape, until the stars are fully exposed. 
Overall, the evolution is very monotonous. At a given radius from the star, eg. 10 pc, the density is smoothly decreasing with time, and the fraction of ionized gas is increasing (left column of Fig.~\ref{fig:HIIRegion}).

When averaging over all O stars at the solar circle (middle panel of the two first rows of Fig.~\ref{fig:HIIRegion}), the same progressive evolution emerges. 
Only less massive O stars ($M \lesssim 20$~\Msun) survive until $10$~Myr, and those have a much lower energy contribution (Fig.\ref{fig:IonisationFraction}), which is why the average ionization fraction is less than one at $10$~Myr. Those low mass O stars are struggling to fully disrupt their high density surrounding. 

\subsection{The case of the CMZ}

In order to know whether the extreme environment of the CMZ still allows for the classical GMC evolution sketched in Fig.~\ref{fig:SkechDisk}, we follow O stars in the CMZ and explore their surrounding.

The CMZ being a much higher surface-density environment, we expect ionizing radiation there to be less efficient at carving out a low density cavity, disrupting the parental cloud from within. In Fig.~\ref{fig:HIIevo} (bottom) we have selected an O star in the CMZ, and an observer at its position indeed experiences higher surface densities, a generally smaller H{\sc ii} region, and smaller windows where the ionizing radiation can escape.

On the other hand, due to the higher star formation rate surface density, we also expect more massive young clusters, whose combined feedback action could disrupt the high-density environment. 
In practice, the short dynamical timescale in the CMZ is comparable to the feedback timescale, meaning that fresh dense molecular gas can easily flow into the star formation site, re-embedding the O star. This is the picture that we see by following the early lifetime of an O star in the CMZ, shown in Fig.~\ref{fig:HIIevo} (bottom). Initially, the strong ionizing radiation from the cluster is able to expose the star, and carve out an H{\sc ii} region. Yet, after some time, owing to the dense and dynamic nature of the surroundings, new molecular material is flowing in, re-embedding the star and closing the windows for radiation to escape. At later stages yet, the star can become re-exposed once more. 

If we look at the density profile and structure of the H{\sc ii} region surrounding this randomly selected O star (Fig.~\ref{fig:HIIRegion}, left panel of the two bottom rows), we can spot this episodic nature as well. The star is born in a high-density environment, is able to disrupt its surrounding within $10$-$20$~pc over their first $\sim 4$~Myr lifetime, only to be re-embedded fully at around $5$-$7$~Myr. At an age of $10$~Myr the stellar particle is exposed again.

Looking at the averaged density and ionization profiles of all stellar particles holding at least one O star in the CMZ (Fig.~\ref{fig:HIIRegion}, middle column), we are not able to see this episodic behavior as the time evolution of each star is different and will average out (see the right column of the same figure). However, we learn that on average, the background density at $50$~pc of H{\sc ii} regions in the CMZ is around $\sim 30$~\cc, much higher than at the solar circle. Moreover, on average the H{\sc ii} region is able to ionize only a small fraction of their surrounding, ionizing at most $\sim 30$~\% of a surface with $r=50$~pc. This is due to the higher surface density environment, but also because of the large fraction of stars undergoing these episodic high-density embedding events. We see that at least on average the star is still able to clear its high density birth environment over its initial $4$-$6$~Myr, but densities increase again and H{\sc ii} regions become more embedded at later times. The reason can be found in the higher cadence of re-embedding events for less massive O stars at later ages in a dynamic high-surface density environment, as the more massive and energetically radiating O stars died off earlier. 

\begin{figure}
	\includegraphics[width=\columnwidth]{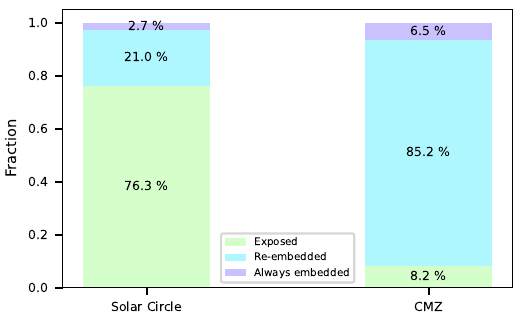}
        \caption{Fraction of stars classified based on their evolution (clarified in the text) for the solar circle and the CMZ. Most of the O stars in the CMZ will undergo at least one event where the star after being exposed, will be re-embedded in a high density environment, in stark contrast to what O stars at the solar circle experience.} 
    \label{fig:StellarTypes}
\end{figure}
To get a more quantitative comparison, we classify stars based on the evolution of their surrounding environment. We use the ionized surface fraction $f_{S_{\rm ion}}(r)=S_{\rm ion}(r)/(4\pi r^2)$ at a distance of $r=10$~pc, where $S_{\rm ion}(r)$ is the surface around the star at a distance $r$ which the star contributes at keeping ionized, i.e. when a ray from the star only passes through ionized gas ($x_{\rm H^+} > 0.5$). The star is categorized as always embedded if the feedback is unable to expose the star, and $f_{S_{\rm ion}}(10 \, \rm pc) < 0.5$ throughout their lifetime up to a maximum of $t=10$~Myr. On the other hand, we classify star particles as exposed if at some point $f_{S_{\rm ion}}(10 \, \rm pc) > 0.5$ and it remains that way for the rest of their lifetime up to $10$~Myr. Finally, we label stellar particles as re-embedded if at some point, after being exposed with $f_{S_{\rm ion}}(10 \, \rm pc) > 0.5$, the star enters a high-density environment and is strongly embedded again, with $f_{S_{\rm ion}}(10 \, \rm pc) < 0.2$.

In Fig.~\ref{fig:StellarTypes} we show the fraction of star particles in each category with at least one O star, for the CMZ and the solar circle separately. We see that most stars in the solar circle end up being exposed without experiencing any re-embedding event, which is consistent with the scenario sketched of Fig.~\ref{fig:SkechDisk} \footnote{The small but significant fraction of stars classified as re-embedded can be explained when early SF is still unable to fully disrupt the molecular complexes, and new dense material is continuing to rush toward the SF site}. 
In contrast, most of the stars in the CMZ experience at least one re-embedding event.
This implies that the classical GMC evolution scenario does not apply in the CMZ and a new model is needed. Only a small fraction of stars remain fully embedded throughout their lifetime in both environments, which is a confirmation that early ionization feedback is effective in generating low-density HII regions. 

\section{Toward a new model for feedback coupling in the CMZ}
\label{sec:discussion}

\subsection{Dynamical decoupling of stars and gas}
\label{subsec:decoupling}

\begin{figure*}
	\includegraphics[width=\textwidth]{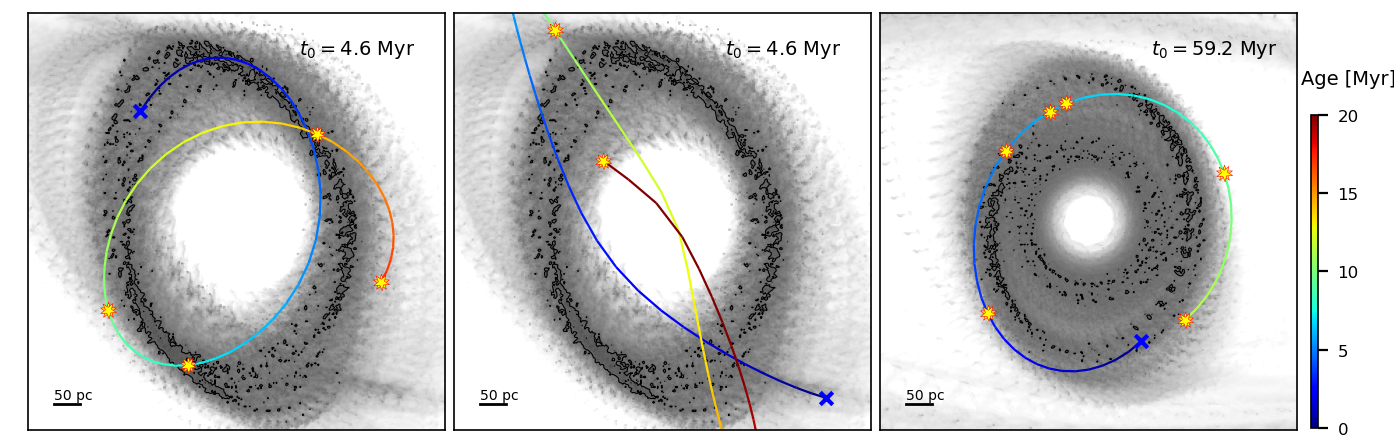}
        \caption{Orbits of a few selected stellar particles containing O stars at birth. The orbit is colored by the age of the stellar particle, and each SN event in its lifetime associated with the particle is marked along its orbit. The birthplace of the star is highlighted with the blue x-marker. In the background in grayscale we show the averaged H$_2$ density over a period of $20$~Myr starting from the snapshot where the star was born. The black contour shows an average density of $50$~\cc. In the first panel we show an early snapshot when the inflow from the bar lanes is still quite copious. In the middle panel we show the orbit of a star which formed in a GMC associated with the bar lanes and inherited their velocity. On the right panel a snapshot at a later time was chosen, where the CMZ is more regular, and the inflow smaller and more steady. Stars quickly decouple from the gas on orbital timescales, which are short in the CMZ.} 
    \label{fig:StellarDecoupling}
\end{figure*}

\begin{figure}
	\includegraphics[width=\columnwidth, trim=.5cm 0.cm 0cm .5cm, clip]{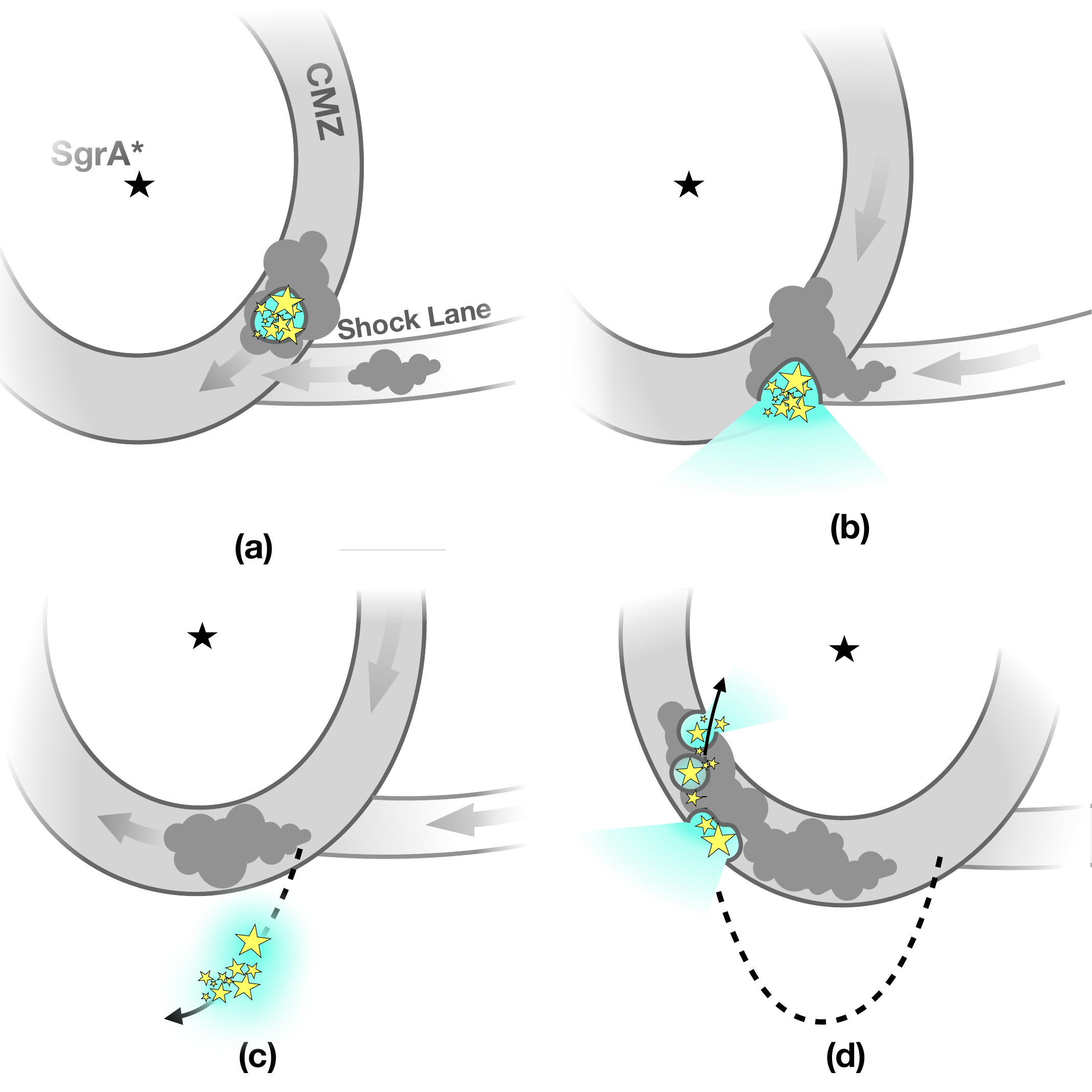}
        \caption{Schematic representation of the early evolution of a star-forming region in the CMZ and the subsequent evolution of the formed stars. Compared to Fig~\ref{fig:SkechDisk}, the situation is more complex in the dynamic environment of the Galactic Center. Stellar-gas decoupling (panel b) due to cloud collisions and short orbital times, leads to re-embedding events in the early evolution of the stars (panel d). This has an effect on how feedback can couple to the dense gas in the CMZ.} 
    \label{fig:SkechCMZ}
\end{figure}
The task now is to understand and describe the causes of the differences in behavior of how these stars affect their surroundings in the two different environments. Just referring to the more extreme conditions of the CMZ is unsatisfactory as an explanation. These certainly contribute in making it harder for the ionizing feedback to efficiently disrupt the surrounding molecular gas, but what makes the CMZ so special is that the gas there follows the very specific dynamics shaped by the barred potential. The gas circles around the Galactic Center on $x_2$ orbits with orbital times of the order of $5$-$10$~Myr, comparable to GMC lifetimes, so this has to be factored in in any description of star formation and cloud evolution in the region. 

One major effect is that gas and stars undergo different physical processes, and while gas follows the equation of hydrodynamics and responds to pressure forces such as gas collisions and shocks, stars are collisionless and only feel gravitational forces. This leads to gas and star decoupling on times comparable to the orbital timescale which in the CMZ are short enough to matter for star formation and GMC evolution. In essence, the gas in the CMZ can only follow nearly closed $x_2$ orbits. Otherwise, it would collide with its own self-intersecting orbit, dissipating energy and falling back on the nearly closed orbit. Stars, on the other hand, will generally follow open rosetta-type orbits \citep{Nieuwmunster2024}, and therefore decouple from the gas on an orbital timescale (Fig.~\ref{fig:StellarDecoupling}). The high-density and turbulent environment exacerbates this behavior, and any cloud collision and shock will enhance the velocity and spatial difference between star and surrounding material from which the star formed. \citet{Tassis2022} estimate the stellar velocity dispersion to be large enough such that even young O stars can cross entire GMCs within their lifetime. Moreover, small differences in velocity and/or position, can quickly grow on short timescales in the environment governed by strong potential gradients, which leads to high orbital angular velocities and strong shear. 
Finally, the CMZ experiences violent and frequent gas accretion events through radial gas-inflow from the bar lanes. These inflow events happen on timescales relevant to star formation and GMC evolution as well, and can play an important role in shaping the dynamics of the CMZ. The new inflowing material can cover up freshly exposed stars and clusters, and colliding with the gas in the CMZ, can enhance gas-star decoupling. 

In periods of more violent accretion, the CMZ can be disrupted and become asymmetric. As a consequence, stellar orbits can be more heated in phase-space, leading to more extreme and frequent decoupling between gas and star. This can be seen in the first panel of Fig.~\ref{fig:StellarDecoupling}, where a stellar orbit is shown with respect to the averaged gas distribution at an early time of the simulation, when the accretion from the bar lanes is quite violent. The star exhibits a noticeable open orbit. At later times, when the inflow is lower and steadier, the decoupling is still present, but less extreme (last panel of Fig.~\ref{fig:StellarDecoupling}). For clarity, we only plot one orbit, but most stars follow a similar behavior. 
The decoupling between gas and stars is taken to the extreme when the star formation is associated with the bar lanes as can be seen in the middle panel of Fig.~\ref{fig:StellarDecoupling}. The gas will collide with CMZ gas, while the star will continue to follow the high-angular momentum $x_1$ orbits, a family of orbits elongated parallel to the bar major axis and much more elongated than the $x_2$ orbits. The H{\sc ii} regions of SgrE \citep{Anderson2020} are an observational example of this behavior. 

In essence, in the CMZ, the dynamical decoupling between star and gas happens on a very short timescale, when the star is still in its feedback stage. 
Because of this decoupling, the behavior of a typical young star-forming region in the CMZ is very different than the classical scenario in the solar circle illustrated in Fig.~\ref{fig:SkechDisk}.
We have sketched out an adapted scenario of the early evolution of a star-forming region in the CMZ in Fig.~\ref{fig:SkechCMZ}, accounting for the decoupling between star and gas which, in turn, triggers the re-embedding events described in section \ref{sec:env_o_stars}.

\subsection{Cluster shearing}
\label{subsec:shearing}

\begin{figure*}
	\includegraphics[width=\textwidth]{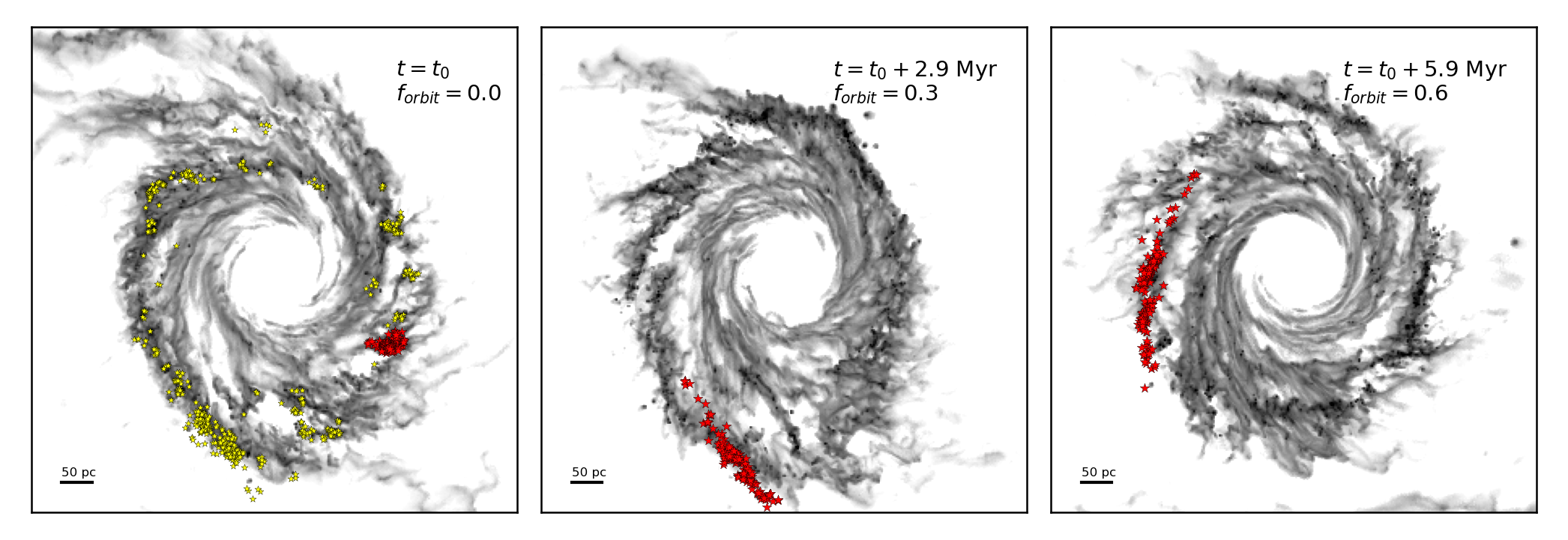}
    \caption{Young stellar association that is quickly sheared apart in the CMZ environment. We highlight new-born ($< 3$~Myr) O stars at a given time in the simulation (left, yellow stars) and select a compact association of coeval O stars in red. We follow these stars in their evolution and show their position with respect to the background molecular gas (in grayscale) after $t=3$~Myr (middle) and $t=6$~Myr (right), which correspond to orbital fractions of $f_{\rm orbit} = 0.3$ and $f_{\rm orbit} = 0.6$ respectively.} 
    \label{fig:clusterShearing}
\end{figure*}

Another major effect on early evolution of star-forming regions and stars is given by the strong shear in the CMZ region compared to the Disk. This has the effect that small initial velocity and position differences, will quickly increase in amplitude on orbital timescales. Coeval stellar associations which formed from the same cloud complex, will be sheared apart much more efficiently and on smaller timescales compared to the solar circle (Fig.~\ref{fig:clusterShearing}). The dissolution of star clusters due to the Galactic tidal field has been estimated to happen on timescales of the order of just a few megayears in the Galactic Center \citep{PortegiesZwart2002, Kruijssen2014}. Indeed, in the CMZ only two young clusters are known (Arches and Quintuplet) whose masses alone cannot account for all the recent star formation in the region. This is known as the missing clusters problem. Moreover, evidence for the presence of young dispersed stellar associations in SgrB1 \citep[$\sim10$~Myr old,][]{Nogueras-Lara2022} and Sgr~C \citep[$\sim20$~Myr old,][]{Nogueras-Lara2024} have been reported. 

In the simulation, this effect generates stellar streams of coeval stars along the simulated CMZ (Fig.~\ref{fig:Projections}, panel of the face-on view of the CMZ where star particles are colored by age). As in the observations, we do see a few surviving clusters of stars in the projected stellar surface density, but the bulk of the young stars are dispersed throughout the CMZ.

\subsection{Implications on the efficiency of stellar feedback}
\begin{figure}
    \centering
    \includegraphics[width=\linewidth]{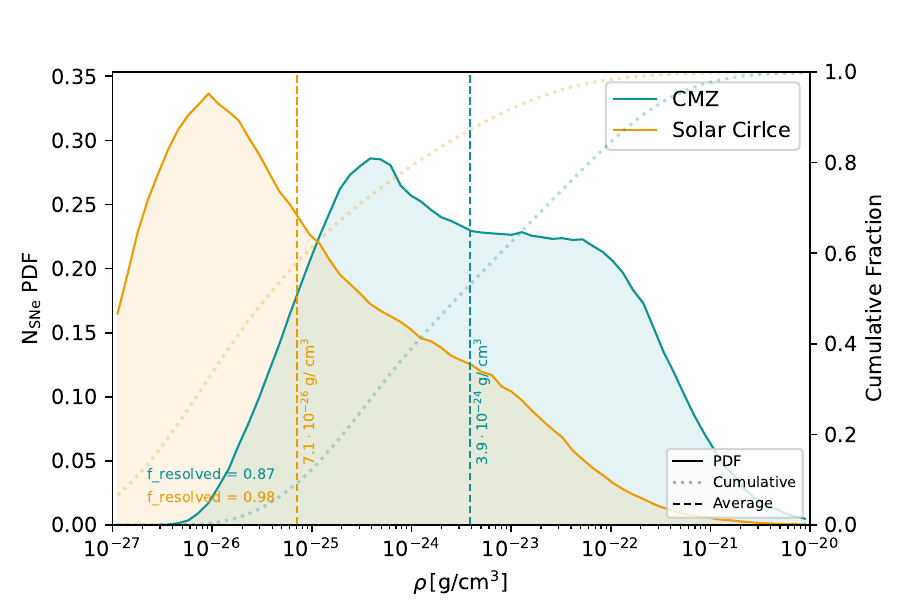}
    \caption{Distribution of the average gas density of the injection regions (defined in Sec.~\ref{sec:supernovae}) of supernova events. The distribution of SNe (solid lines) and the cumulative distributions (dotted lines) are shown. SNe in the CMZ and the solar circle are depicted in different colors. The dashed vertical lines indicate averages of the distributions. In the bottom left we also note the fraction of resolved SN events, for which we can inject thermal energy as opposed to kinetic energy. }
    \label{fig:SNe-density}
\end{figure}

Both the dynamical decoupling of the gas and stars (section \ref{subsec:decoupling}) and the shearing of the stellar clusters (section \ref{subsec:shearing}) have a noticeable impact on how the feedback couples back to the surrounding ISM and affect how efficiently clouds are disrupted. 

Classically, in the disk, ionizing radiation and SNe will disrupt the GMC from which the star was formed. However, in the CMZ, stellar decoupling and cluster shearing have the consequence that the feedback is injected into gas which is dynamically de-correlated with the star. The dense gas, left without ionizing stars, may continue its collapse and form new stars, such that these processes have an impact on the star formation efficiency of GMCs. By contrast, dense gas in other parts of the CMZ may host wandering O stars and be disrupted before being able to collapse. \citet{Tassis2022} estimate that such wandering O stars could explain the high level of turbulence as well as account for the steep size-linewidth relation in CMZ clouds.

Additionally, cluster shearing will preclude the compounding effect of clustered feedback, which is an important aspect for cloud disruption and feedback coupling \citep[e.g.][]{Walch2015,Gentry2019}. Indeed, unless stars are formed in a sufficiently massive and compact bound cluster, more extended stellar associations will not survive long in their $x_2$ orbit. As a consequence, the feedback from the stars is less concentrated, and instead more spread-out over the entire CMZ and thus less efficient in disrupting the surrounding molecular gas. These stars will act as a more evenly distributed background feedback source, driving turbulence, without being able to create large superbubbles. In isolation, they are then much more easily re-embedded in the turbulent flow of the CMZ. 

These special CMZ conditions will have an effect on the environment into which supernovae are injected. In the Disk, they are predominantly injected at the center of superbubbles, in progressively lower density. In the CMZ, instead, due to the previously mentioned effects, they will be more randomly distributed, such that SNe in higher density gas are more frequent (Fig.~\ref{fig:SNe-density}). This might change the way in which stellar feedback impacts the ISM in the CMZ as a whole, for instance in driving turbulence \citep{Iffrig2015,Walch2015SN,Ohlin2019}.

This could be another puzzle piece in explaining how SF is regulated in the Galactic Center. Contrasting theories, each supported by simulations, state that either the SF in the CMZ is highly episodic \citep{Krumholz2017, Armillotta2019, Nogueras-Lara2020}, with phases of intense SF followed by phases where the feedback disrupts the CMZ entirely and shuts SF off. Or conversely, SF is rather constant in time, and regulated by the inflow. The solution to the conundrum reconciling the two hypotheses could lie in how stars are formed in the CMZ. In one case stars form in more dense and compact clusters, surviving the strong shearing environment and the resulting feedback-driven superbubbles are much more disruptive to the gas in the CMZ. In the other case, less compact clusters form and get easily sheared apart. The stellar feedback then provides a background source of turbulence \citep{Tassis2022}, but is unable to fully shut SF off, as the high-density gas is inefficiently being disrupted, and instead the amount of dense gas is regulated primarily by the inflow. 

Stellar-gas dynamical decoupling, and cluster shearing strongly depend on the properties of the local potential, orbital times and size of the CMZ. As the CMZ is believed to be even smaller in observations compared to our simulations, it is plausible that these effects are even more strongly pronounced in the real CMZ. Observations of external CMZs \citep[e.g.][]{Schinnerer2023, Sun2024} could be used to corroborate this hypothesis. 

\section{Caveats}
\label{sec:caveats}

\begin{figure}
	\includegraphics[width=\columnwidth]{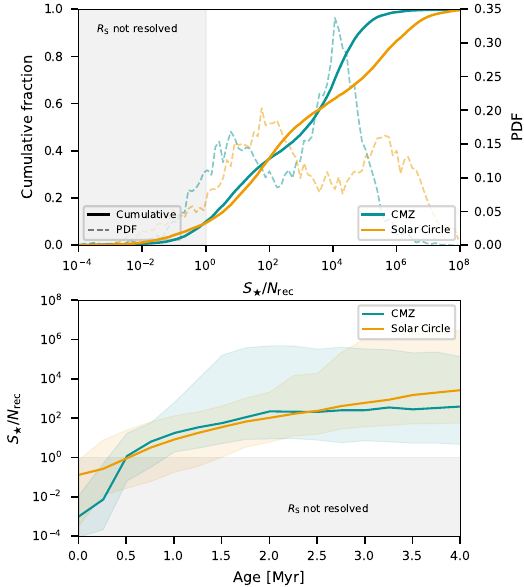}
        \caption{Distribution of the ratio of the ionizing photon flux $S_\star$ of star particles containing at least one O star over the recombination rate of its closest gas cell $N_{\rm rec}$. This is a measure of how well the Str\"{o}mgren radius is resolved. In the top panel, we show the (cumulative) distribution of all stellar particles containing at least one O star for a given snapshot. For the bottom panel, we follow the evolution of individual star particles and show $S_\star/N_{\rm rec}$ averaged over all star particles as a function of their age. The shaded region shows $\pm \sigma$ around that average.}
    \label{fig:StromgrenResolution}
\end{figure}

\subsection{Resolution of the H{\sc ii} regions}
There are a few limitations of these simulations which could quantitatively affect our results. 
The main limitation is that we are not able to resolve the most compact and embedded initial phase of the evolution of H{\sc ii} regions around O stars.

Indeed, the dotted lines in the top panel of Fig.~\ref{fig:Resolution} mark the size of a Str\"{o}mgren sphere as a function of density and are demarcation lines for when the radiation feedback is resolved, based on the mass of the star. Note the steepness of these lines, which shows how expensive it would be to require the Str\"{o}mgren radius to always be resolved. On the other hand, at later stages of the evolution of young massive stars, when the density progressively decreases, the interaction between the stellar radiation and the surrounding ISM is captured with increasing accuracy. The situation is even more hopeless if we want to resolve the radiation coming from B stars as well. Luckily most of the energy in ionizing radiation comes from the more massive O stars (see Fig.~\ref{fig:IonisationFraction}). 

Since star particles are considered a small association of stars, more than one OB star might be contributing to the radiation feedback of a single particle, helping us resolving the surrounding Str\"{o}mgren sphere. All this considered, we can see from Fig.~\ref{fig:StromgrenResolution} that while the early stages of the H{\sc ii} region are not resolved throughout the simulation, after just $0.5$-$1$~Myr the stellar particles enter density regimes where, on average, excellent resolutions are achieved. Moreover, at any point in time, only $\sim 10$~\% of O stars in the domain are in a regime where the Str\"{o}mgren sphere is unresolved.
Better capturing this initial evolution could lead to more efficient disruption of the surrounding gas, affecting timescales of the evolution.

\subsection{Softening of the gravity solver}

Decoupling times and stellar dynamics would change if an unsoftened gravity solver was used and all the complexity of stellar interactions in associations and clusters would have been considered (such as binaries, close encounters etc...). In particular, stellar clusters are more easily dynamically dissolved in less resolved N-body simulations \citep{Dehnen2001, Heggie2003}. On the other hand, if the formed cluster is tightly bound, gravitational softening does not allow for strong gravity kicks from close stellar encounters, and the cluster stays compact \citep[e.g.][their figure 10]{Lahen2025}. However, given our coherent comparison of the two environments within the same simulation framework, we can robustly highlight qualitative differences arising from the different conditions experienced by the stars and gas. 

\section{Summary and conclusions}
\label{sec:conclusion}

We described a radiation-MHD simulation of a truncated Milky Way galaxy. We focused on and compared two high-resolution regions, a ring at the solar radius and the inner Galaxy, to understand and describe how the special environment of the CMZ affects ISM properties, star formation and early evolution, and feedback coupling.

Our main findings are summarized below.
\begin{itemize}
    \item The CMZ is a much harsher environment than the disk. This affects the ISM properties and SF. The gas in the CMZ is denser on average, and a larger mass fraction is cold and molecular. The magnetic fields in the CMZ are enhanced in every density regime by up to one order of magnitude. The ISM is more turbulent with $\sigma_z\sim9$~km/s against $\sigma_z\sim2$~km/s for the solar circle. The star formation rate surface densities are higher, but the molecular depletion times are similar. However, at comparable molecular gas surface densities, the CMZ seems to be less efficient in converting gas into stars than in the solar circle.
    \item The CMZ is not just a high surface density analog of the disk (inflow, short orbital times, etc. ). This leads to a difference in the interaction of young massive stars with their environment. In the disk, O stars disrupt clouds from within, following a progressive evolution, from dense and highly embedded to progressively more diffuse and exposed. Instead, O stars in the CMZ mostly follow cycles where the star once being in an exposed state, will be re-embedded at some later point.
    \item The difference in behavior can be explained by the high surface density environment, by the highly turbulent and  dynamic ISM, by frequent cloud collisions between GMCs within the CMZ, and with inflowing gas from the bar lanes, but most noticeably, it can be explained by quick decoupling of stars and gas in the CMZ and by the efficient dynamical disruption of stellar associations through shear.
    \item These two effects strongly determine the coupling of feedback to the (dense) ISM in the CMZ and the regulation of SF. Dynamical decoupling and cluster shearing cause the feedback to be deposited far away from the GMC structure in which the stars formed. Moreover, the shearing deprives the ISM of the combined effect of clustered feedback and its disrupting power through strong superbubbles. Therefore, feedback is less disruptive for the global CMZ structure, but instead acts as a background source of turbulence. 
\end{itemize}

\begin{acknowledgements}
We thank the anonymous referee for constructive feedback on the manuscript. The authors are grateful for computing resources provided by the Ministry of Science, Research and the Arts (MWK) of the State of Baden-W\"{u}rttemberg through bwHPC and the German Science Foundation (DFG) through grants INST 35/1134-1 FUGG and 35/1597-1 FUGG, and also for data storage at SDS@hd funded through grants INST 35/1314-1 FUGG and INST 35/1503-1 FUGG. VMP, NB, RSK, SCOG, SR, JS, JG, and PG gratefully acknowledge financial support from the European Research Council via the ERC Synergy Grant "ECOGAL" (project ID 855130). FNL gratefully acknowledges financial support from grant PID2024-162148NA-I00, funded by MCIN/AEI/10.13039/501100011033 and the European Regional Development Fund (ERDF) “A way of making Europe”, from the Ramón y Cajal programme (RYC2023-044924-I) funded by MCIN/AEI/10.13039/501100011033 and FSE+, and from the Severo Ochoa grant CEX2021-001131-S, funded by MCIN/AEI/10.13039/501100011033. MCS, MD and ZXF acknowledge financial support from the European Research Council under the ERC Starting Grant ``GalFlow'' (grant 101116226). MCS further acknowledges support from Fondazione Cariplo under the grant ERC attrattivit\`{a} n. 2023-3014.
NB acknowledges support from the ANR BRIDGES grant (ANR-23-CE31-0005).
RSK and TP also acknowledge support  from the German Excellence Strategy via the Heidelberg Cluster ``STRUCTURES'' (EXC 2181 - 390900948). In addition RSK is grateful for funding from the German Ministry for Economic Affairs and Climate Action in project ``MAINN'' (funding ID 50OO2206), and from DFG and ANR for project ``STARCLUSTERS'' (funding ID KL 1358/22-1). 
MH, JP and AP acknowledge funding from the Swiss National Science Foundation (SNF) via a PRIMA grant PR00P2 193577 `From cosmic dawn to high noon: the role of BHs for young galaxies'.
\end{acknowledgements}

\bibliographystyle{aa}
\bibliography{bibliography}

\end{document}